\documentclass[10pt]{article}

\usepackage{epsfig}
\usepackage{natbib}
\usepackage{url}
\usepackage{amsmath}
\usepackage{dsfont}
\usepackage{graphicx,graphics}
\usepackage{amssymb,amsmath}
\usepackage{verbatim,enumerate}
\usepackage{rotating}
\usepackage{multicol,color}
\usepackage{lmodern}
\usepackage{array}
\usepackage{multirow}
\usepackage{tabularx}
\usepackage{lscape, pdflscape}
\usepackage{url}
\usepackage{dcolumn}
\usepackage{times}
\usepackage{setspace}
\RequirePackage{lineno}

\def\log{\hbox{log}}

\def\boxit#1{\vbox{\hrule\hbox{\vrule\kern6pt
          \vbox{\kern6pt#1\kern6pt}\kern6pt\vrule}\hrule}}
\def\refhg{\hangindent=20pt\hangafter=1}
\def\refmark{\par\vskip 2mm\noindent\refhg}

\def\refhg{\hangindent=20pt\hangafter=1}
\def\refmark{\par\vskip 2mm\noindent\refhg}

\def\bse{\begin{eqnarray*}}
\def\ese{\end{eqnarray*}}
\def\be{\begin{eqnarray}}
\def\ee{\end{eqnarray}}
\def\bq{\begin{equation}}
\def\eq{\end{equation}}
\def\bse{\begin{eqnarray*}}
\def\ese{\end{eqnarray*}}

\newtheorem{Th}{Theorem}

\def\btheta{{\boldsymbol \theta}}

\def\part{\partial}

\newcommand{\bd}{{\bf d}}

\begin{document}

{\center\LARGE  Growth curve based on scale mixtures of skew-normal distributions to model the age-length relationship
of Cardinalfish ({\it Epigonus  Crassicaudus})\\}
\vskip10mm

{\center {\bf Javier E. Contreras-Reyes}\\
\vskip2mm
Divisi\'on de Investigaci\'on Pesquera, Instituto de Fomento Pesquero\\

Av. Almirante Blanco Encalada 839, Valpara\'iso, 2361827, Chile\\

Email: jecontrr@mat.puc.cl\\}

\vskip10mm

{\center {\bf Reinaldo B. Arellano-Valle}\\
\vskip2mm
Departamento de Estad\'istica, Pontificia Universidad Cat\'olica de Chile\\

Santiago, Chile. Email: reivalle@mat.puc.cl\\}
\vskip10mm

\begin{abstract}
Our article presents a robust and flexible statistical modeling for the growth curve associated to the age-length
relationship of Cardinalfish ({\it Epigonus Crassicaudus}). Specifically, we consider a non-linear regression model,
in which the error distribution allows heteroscedasticity and belongs to the family of scale mixture of the skew-normal
(SMSN) distributions, thus eliminating the need to transform the dependent variable into many data sets. The SMSN is a
tractable and flexible class of asymmetric heavy-tailed distributions that are useful for robust inference when the
normality assumption for error distribution is questionable. Two well-known important members of this class are
the proper skew-normal and skew-$t$ distributions. In this work emphasis is given to the skew-$t$ model.
However, the proposed methodology can be adapted for each of the SMSN models with some basic changes. The present
work is motivated by previous analysis about of Cardinalfish age, in which a maximum age of 15 years has been
determined. Therefore, in this study we carry out the mentioned methodology over a data set that include a long-range
of ages based on an otolith sample where the determined longevity is higher than 54 years.\\

\end{abstract}

\vskip15mm
\section{Introduction}\label{sec.intro}


Currently, an increasing interest in describing the growth of biological species can be found amongst different studies.
This has motivated the use of some biological models proposed in the literature to describe the growth associated with
the age-length relationship. Among these works, the von Bertalannfy (VB) growth curve can be found (von Bertalanffy,
1938; see also Allen, 1966; Kimura, 1980; Gamito, 1998), which explains the length of a specie in terms of its age
by means of a non-linear function depending on tree parameters $(L_{\infty},K,t_0)$, where $L_{\infty}$ represents
the asymptotic length of the specie under study, $K$ is the growth rate (also known as the Brody growth coefficient of)
and $t_0$ is the theoretical age at the zero length. Specifically, if $y$ represents the observed length at age $x$,
then a deterministic expression of the VB growth curve is given by
\begin{equation}\label{VB}
y=L_{\infty}(1-e^{-K(x-t_0)}).
\end{equation}
In order to fit the equation (\ref{VB}) from an empirical dataset, $(y_t,x_t)$, $t=1,...,n$ say, where $y_t$ (length)
and $x_t$ (age) are the response and explanatory variables, respectively, the VB growth curve can be described in terms
of a non-linear regression as
\begin{equation}\label{NL}
y_t=\eta_t+\varepsilon_t,
\end{equation}
$t=1,...,n$, where $\eta_t=\eta(\beta;x_t)=L_{\infty}(1-e^{-K(x_t-t_0)})$, $\beta=(L_{\infty},K,t_0)^T$ is the vector of unknown parameters and the $\varepsilon_t$ are independent random errors.

Kimura (1980) studied the relation (\ref{NL}) under the assumptions of independence an normality for the random errors models, $\varepsilon_t\buildrel ind.\over\sim N(0,\sigma^2)$ say, and proposed the maximum likelihood method to fit the model (see also Allen, 1966). More recently, Cubillos et al. (2009) studied the VB model however using the Cope \& Punt methodology (Cope \& Punt, 2007), which considers a random error in assigning age, which is determined by two different readers. Although this last model also considers the independence and normality assumptions for the error terms, it assumes that the assigned age is determined by an exponential or gamma distribution, in this way guaranteeing a real age composition. In addition, to empirically studying the
age-length relationship, Cubillos et al. (2009) considered samples of otolith of Cardinalfish obtained from 1998 to 2007 in the Chilean south central coastal zone (Latitude $33^{\circ}$S - $42^{\circ}$S), from which a random selection of 96 otolith was obtained within the range of 20-37cm, and with ages of less than 15 years (G\'alvez et al., 2000). However, a new method to read otolith which is described in detail by Ojeda et al. (2010), formulates the hypothesis that Cardinalfish could have a longevity of at least of 54 years. Furthermore, this specie is characterized by living in waters from 100 to 550m in depth, but generally between depths of 250 and 300m; and according to commercial capturing registers, lengths varying mostly between 17 and 47cm, which does not present significative differences in either sex (Wiff et al., 2005).

In this paper, we study the VB growth model (\ref{VB}) considering a flexible class of non-normal distributions for
the random error $\varepsilon_t$. Specifically, as in Basso et al. (2010), we consider the class
of scale mixture of the skew-normal (SMSN) distributions (Branco \& Dey, 2001) for random errors.
The SMSN is an attractive class of asymmetric heavy-tailed distributions that are useful for robust inference when
the normality assumption for error distribution is unrealistic. Some important members of this class are the skew-t, skew-slash, and
skew-contaminated normal distributions (see, e.g., Lachos et al., 2010). The flexibility of these distributions allow to fit observations with a high presence of skewness and heavy tails, and they are useful to model some aleatory phenomenon with extreme values which generate residual heterogeneity in classical models (Kimura, 1990). To estimate the parameters of the SMSN models, Labra et al. (2012) implemented the expected conditional maximum estimation (ECME) algorithm of Liu \& Rubin (1994). They incorporated a correction to the EM algorithm to estimate the parameters of a non-linear model
in a fast and more robust form.  In addition, Labra et al. (2012) implemented some model comparison and diagnostic methods. Our study considers
the local influence analysis of Cook (1986) and of Poon \& Poon (1999). Our results are presented considering a dataset containing observations
of species of at least 61 years old, so that it has a larger range of variability than that of the G\'alvez et al. (2000) study. Hence, the use of this dataset could strongly affect the estimations of parameters $L_{\infty}$, $K$ and $t_0$.

This paper is organized as follows. Section 2 presents a description of the main
theoretical aspects of the methodology implemented by Labra et al. (2012). Section 3 shows the empirical behavior of lengths
across different age categories without distinction of genre. This section also includes the main estimation and
diagnostic results. Finally, Section 4 concludes with a discussion of these results.\\

\section{Methodology}

In this section, we study the VB non-linear regression model using a similar approach to that used by Lachos et al. (2011), Labra et al. (2012) and Basso et al. (2010). Specifically, we consider the non-linear regression model (\ref{NL}) with the assumption that the random errors $\varepsilon_t$ are independent, heteroscedastic and distributed throughout SMSN class of distributions.
In other words, we suppose that
\begin{equation}\label{E-SR}
\varepsilon_t=v_t^{-1/2} e_t+\mu_t,
\end{equation}
$t=1,\ldots,n$, where the $e_t$ and $v_t$ are independent random quantities and the $\mu_t$ are location parameters. More precisely, the $e_t$ are independent and heteroscedastic skew-normal random errors, $e_t\buildrel ind.\over\sim SN(0,\sigma_t^2,\lambda_t)$ say, i.e., with density function
$$h(e_t;\sigma_t,\lambda_t)=\frac{2}{\sigma_t}\phi\left(\frac{e_t}{\sigma_t}\right)\Phi\left(\lambda_t\frac{e_t}{\sigma_t}\right),\quad -\infty<e_t<\infty,$$
where $\sigma_t>0$ and $-\infty<\lambda_t<\infty$ are scale and shape/skewness parameters, respectively, and $\phi(z)$ and $\Phi(z)$ are, respectively, the density and distribution function of the standardized normal distribution. Meanwhile,  in (\ref{E-SR}) the $v_t$ are positive (scale) random factors  perturbing the skew-normality, which are assumed to be independent and identically distributed (iid) with distribution function $G(v;\nu)$ defined on $(0,\infty)$ and depending on the unknown parameter $\nu$ (possibly vectorial).

As we know, the mean and variance of the skew-normal random errors $e_t\sim SN(0,\sigma_t^2,\lambda_t)$ are given by
$E(e_t)=\sqrt{2/\pi}\,\delta_t\sigma_t$ and ${\rm Var}(e_t)=\left\{1-\left(2/\pi\right)\delta_t^2\right\}\sigma_t^2$,
where $\delta_t=\lambda_t/\sqrt{1-\lambda_t^2}$.
Thus, if we assume in (\ref{E-SR}) that the moments $\kappa_{k}=E(v_t^{-k/2})$, $k=1,2$, are finite, then the mean and variance of the SMSN random errors  $\varepsilon_t$ exist which are given by
$E(\varepsilon_t)=\sqrt{2/\pi}\,\kappa_1\sigma_t\delta_t+\mu_t$ and ${\rm Var}(\varepsilon_t)=\kappa_{2}\sigma_t^2\left\{1-\left(2/\pi\right)\delta_t^2\right\}$.
In order to have errors with zero mean, we impose the condition $\mu_t=-\sqrt{2/\pi}\,\kappa_1\sigma_t\delta_t$. Under this condition, we then have
for the mean and the variance of the response variable $y_t$ that
\begin{equation}\label{r-mean-var}
E(y_t)=\eta_t\quad{\rm and}\quad {\rm Var}(y_t)=\kappa_2\sigma_t^2\left\{1-\left(\frac{2}{\pi}\right)\delta_t^2\right\},
\end{equation}
where $\eta_t=\eta(\beta;x_t)$ is the VB curve defined as in (\ref{NL}).

On the other hand, we can also observe from (\ref{E-SR}) that given the scale mixture factors $v_t$, the random errors $\varepsilon_t$ have skew-normal distribution $SN(\mu_t,v_t^{-1}\sigma_t^2,\lambda_t)$ which are independent. Hence, we have from (\ref{NL}) that conditionally on the $v_t$, the response variables $y_t$ have a distribution given by $y_t\mid v_t\buildrel ind.\over\sim SN(\eta_t+\mu_t,v_t^{-1}\sigma_t,\lambda_t)$, $t=1,\ldots,n$, i.e., the marginal density of  $y_t$ is
\begin{equation}\label{SM/SN-density}
f(y_t;\beta,\sigma_t^2,\lambda_t,\nu)=\frac{2}{\sigma_t}\int_{0}^\infty \sqrt{v_t}\,\phi\left(\sqrt{v_t}\,z_t\right)\Phi\left(\sqrt{v_t}\,\lambda_tz_t\right)dG(v_t;\nu) ,
\end{equation}
$t=1,...,n$, where  $z_t=(y_t-\eta_t-\mu_t)/\sigma_t$, with  $\eta_t=\eta(\beta;x_t)$,  $\mu_t=-\sqrt{2/\pi}\,\kappa_1\sigma_t\delta_t$. 
In this work, we also assume that $\lambda_t=\lambda$, so that $\delta_t=\delta$, and $\sigma_t^2=\sigma^2 m(\rho;x_t)$, where $m(\rho;x_t)$ is a nonnegative function such that $m(0;x_t)=1$. More precisely, we consider the function $m(\rho;x_t)=x_t^\rho$. Consequently, in (\ref{SM/SN-density}) the model parameters are given by $\beta=(L_\infty,K,t_0)^\top,$ $\sigma^2,$ $\lambda$, $\rho$ and $\nu$.

The SMSN class of densities in (\ref{SM/SN-density}) provides different asymmetric heavy-tailed models which are useful to obtain robust inference in the presence of influence observations or outliers. The more well-known densities obtained from (\ref{SM/SN-density}) are the  skew-$t$, skew-slash and skew-contaminated normal (for more details, see Lachos et al., 2010). All these distributions contain the skew-normal one as special case.  Also, for $\lambda_t=0$, the SMSN class reduces to the symmetric class of scale mixtures of normal distributions considered in Lange \& Sinsheimer (1993).\\

\subsection{The skew-$t$ special case}

In this section, we focus our attention principally on the skew-$t$ model with $\nu$ ($\nu>0$) degrees of freedom (Branco and Dey, 2001; Azzalini and Capitanio, 2003; Arellano-Valle et al., 2012). This model follows by assuming in (\ref{E-SR}) that the mixing random factors  $v_t$ are iid $Gamma(\nu/2,\nu/2)$, i.e., with density given by

$$g(v_t;\nu)=\frac{(\nu/2)^{\nu/2}}{\Gamma(\nu/2)}v_t^{\nu/2-1}e^{-\nu v_t/2},\quad v_t>0.$$

In this case, we have $\kappa_1=\sqrt{\nu/2}\Gamma[(\nu-1)/2]/\Gamma(\nu/2)$, $\nu>1$, and 
$\kappa_2=(\nu/2)\Gamma[(\nu-2)/2]/\Gamma(\nu/2)=\nu/(\nu-2)$, $\nu>2$. Also, in (\ref{SM/SN-density}) we obtain
the following skew-$t$ density for the response variables $y_t$:
\begin{equation}\label{ST-model}
f(y_t;\beta,\sigma_t^2,\lambda_t,\nu)
=\frac{2}{\sigma_t}\,t(z_t;\nu)
T\left(\lambda_tz_t\sqrt{\frac{\nu+1}{\nu+z_t^2}};\nu+1\right),
\end{equation}
with $z_t$, $\sigma_t^2$ and $\lambda_t$ defined as in (\ref{SM/SN-density}),  and where
$$t(z;\nu)=\frac{\Gamma[(\nu+1)/2]}{\Gamma(\nu/2)\sqrt{\pi\nu}}\left(1+\frac{z^2}{\nu}\right)^{-(\nu+1)/2},\quad -\infty<z<\infty,$$
i.e., the symmetric Student-$t$ density with $\nu$ degrees of freedom, and $T(z;\nu)$ denotes the corresponding Student-$t$ distribution function. The skew-$t$ (ST) contains the student-$t$ (T), skew-normal (SN) and normal (N) distributions as special cases, as is indicated in the following scheme
\begin{eqnarray*}
\mbox{ST}&\mathop{\longrightarrow}\limits_{\nu\rightarrow\infty}&\mbox{SN}\,\,\mathop{\longrightarrow}\limits_{\lambda=0}\,\,\mbox{N}\\
\mbox{ST}&\mathop{\longrightarrow}\limits_{\lambda=0}&\mbox{T}\,\,\mathop{\longrightarrow}\limits_{\nu\rightarrow\infty}\,\,\mbox{N}\\
\mbox{ST}&\mathop{\longrightarrow}\limits_{\nu\rightarrow\infty,\,\lambda=0}&\mbox{N}\\
\end{eqnarray*}

Let $\theta=(\beta^\top,\sigma^2,\rho,\lambda)^\top=(L_{\infty},K,t_0,\sigma^2,\rho,\lambda)^\top$. From (\ref{ST-model}), and considering the degrees of freedom parameter $\nu$ known, the ST log-likelihood function for $\theta$ is thus given by
\begin{equation}\label{Log}
\ell(\theta)=\sum_{t=1}^{n}\ell_t(\theta),
\end{equation}
where
\begin{eqnarray*}
\ell_t(\theta)&=&\log\mbox{ }2-\frac{1}{2}\log\mbox{ }\sigma^2-\frac{1}{2}\log\mbox{ }x_t^{\rho} +\,\log\mbox{ }\Gamma\left(\frac{\nu+1}{2}\right)-\,\log\mbox{ }\Gamma\left(\frac{\nu}{2}\right)\\
&&-\left(\frac{\nu}{2}\right)\,\log\mbox{ }(\pi\nu)-\left(\frac{\nu+1}{2}\right)\,\log\mbox{ }\left(1+\frac{z_t^2}{\nu}\right)\\
&&+\,\log\mbox{ }T\left(\lambda z_t\sqrt{\frac{\nu+1}{\nu+z_t^2}};\nu+1\right),\\
\end{eqnarray*}
with $z_t=(y_t-\eta_t-\mu_t)^2/\sigma_t$, $\eta_t=L_{\infty}(1-e^{-K(t-t_0)})$, $\mu_t=\sqrt{2/\pi}\,\kappa_1\sigma_t\delta$ and $\sigma_t^2=\sigma^2 x_t^\rho$.

For $\nu$ unknown, an approximation to the maximum likelihood estimator can be computed by varying $\nu$ in a grid of values and
considering the value in the grid that maximizes the likelihood function as an estimator of $\nu$  (Lange \& Sinsheimer, 1993).

The first and second derivatives of $\ell_t(\theta)$ are given in  Basso et al. (2010) for arbitrary specifications of $\eta_t$ and $\sigma_t^2$. From these results we can obtain the observed information matrix, namely
$$J(\theta)=-\sum_{t=1}^{n}J_t(\theta),$$
where $J_t(\theta)=\partial^2\ell_t(\theta)/\partial\theta\partial\theta^{\top}$. Hence, the covariance matrix
of the MLE $\widehat{\theta}$ of $\theta$ can be estimated by $J(\widehat{\theta})^{-1}$, and the respective standard
errors for the components of $\widehat{\theta}$ by diag$\{J(\theta)\}^{-1/2}$. Also, asymptotic confidence interval and
hypothesis testing can be obtained assuming that  $\widehat{\theta}\sim N(\theta,J(\theta)^{-1})$.\\

\subsection{Influence diagnostic analysis}\label{IL}

Influence diagnostic techniques are used to detect observations that may produce excessive
influence in the parameter estimates. There are two main approaches for such techniques:
global influence, which is usually based on case deletion; and local influence, which
introduces small perturbations in different components of the model.

In this work, we consider the local influence analysis proposed by Cook (1986)
to detect observations that exert great influence on the maximum likelihood estimators.
Thus, we focalize our attention in the case-deletion or case-weight approach, in which the impact
of deleting an observation on the estimators is assessed by means of the so called
{\it likelihood displacement} defined by $$LD(\omega)=2\{\ell(\widehat{\theta})-\ell(\widehat{\theta}_{\omega})\}.$$ Here,
$\widehat{\theta}$ and $\widehat{\theta}_{\omega}$ denote the MLEs of $\theta$ under the unperturbed and
perturbed models, respectively, and   $\omega$ is a vector which represents the perturbation scheme, for example, a
collection of case weights. For a given $\omega_0$, we have $l(\theta_{\omega_0})=l(\theta)$ and so  $\widehat{\theta}_{\omega_0}=\widehat{\theta}$.
In this sense, a graph of $LD(\omega)$ versus $\omega$ contains essential information on the influence of the perturbation scheme in question.
Cook (1986) called the geometric surface
$\psi(\omega)=[\omega^{\top},LD({\omega})]^{\top}$ as {\it influence graph}. Also, to characterize the behavior of an
influence graph around $\omega_0$, Cook (1986) defined the normal curvature of
$\psi(\omega)$ in the direction of a vector $d$ of unit length as
$$C_{d}=2\,|d^{\top}H^{\top}J^{-1}H d|,$$
where $J=J(\theta)=-\partial^2\ell(\theta)/\partial\theta\partial\theta^{\top}$
is the observed information matrix and $H=\partial^2\ell(\theta_{\omega})/\partial\theta\partial\omega^{\top}$, which are evaluated at
$\theta=\widehat{\theta}$ and $\omega=\omega_0$. The maximum curvature occurs in the direction of $d_{\max}$,
the eigenvector associated to the largest eigenvalue of the matrix $F=H^{\top}J^{-1}H$. Hence, the vector
 $d_{\max}$ gives information on the direction that $LD(\omega)$ shows  more sensitivity.

Since $C_{d}$ is not invariant under a uniform change of scale,
Poon and Poon (1999) proposed the conformal normal curvature  given by
$$B_{d}=\frac{C_{d}}{2\,\mbox{tr}(F^\top F)^{1/2}},$$
which is such that $0\leq B_{d}\leq1$ for any direction $d$. Thus, they propose to classify the $t$th observation as a possible influential observation if $B_{d_t}$ is greater than the benchmark $c_{d}=\overline{M}_{0}+\tau \sqrt{\mbox{Var}[\overline{M}_0]}$
for a selected constant $\tau$ depending on the observations, and
\begin{eqnarray*}
\overline{M}_0&=&\frac{1}{n}\sum_{t=1}^{n}B_{d_t},\\
\mbox{Var}[\overline{M}_0]&=&\frac{1}{n-1}\sum_{t=1}^{n}(B_{d_t}-\overline{M}_{0})^2.\\
\end{eqnarray*}
This should be interpreted as the effect of the $t$th deleted observation on the log-likelihood function.

\section{Results}

In recent years, several ages estimations for fish living in deep waters have been reevaluated and in many cases maximum ages
have been predicted to be drastically older than those previously considered (Cailliet \& Andrews, 2008). This situation has
been observed in Cardinalfish, where a previous age allocation process, by using the entire otolith sagitta, gives a maximum
age of 15 years (G\'alvez et al., 2000; Cubillos et al., 2009). However, from a new analysis considering the transversal
sections of these otolith, it was found that the Cardinalfish longevity is that of 54 years (Ojeda et al., 2010). Also, the
longevity or life extension (average time between birth and death) are important variables that must be considered in
controlling the exploitation of some species, since  the longevity and growth rate are related directly to the natural
mortality and productivity of the population of those species (Hewitt \& Hoening 2005).

\begin{figure}[htb]
  \centering
  \includegraphics[scale=0.5]{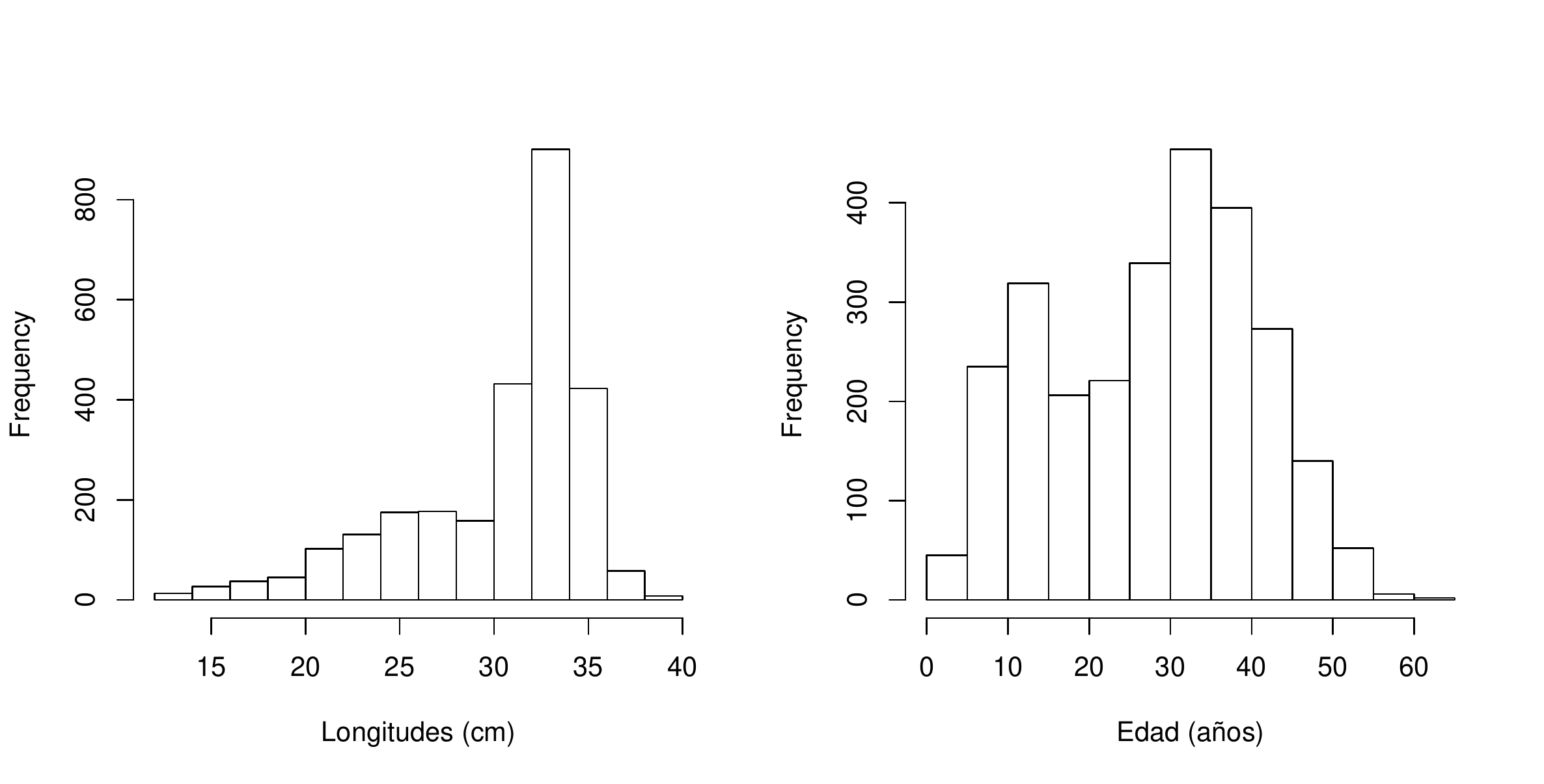}
  \caption{Histograms of ages and lengths for Cardinalfish.}\label{hist} 
\end{figure}


\begin{table}[htb]
\caption{Summary of descriptive statistics for Cardinalfish lengths and age categories.}\label{AD}
\vskip 5 mm
\begin{center}
\begin{tabular}{ccccccc}
  \hline
  Ages & Min. & Max. & Mean & S.D & N. obs. & Proportion\\
  \hline
1 - 3  &  13 & 13 & 13 & - & 1 & 0.04\%\\
3 - 8  &  12 & 26 &19.218 &3.375 &165 & 6.14\%\\
8 - 13  & 19 & 32 &24.317 &2.293 &309 & 11.50\%\\
13 - 18 & 21 & 34 &27.069 &2.374 &261 & 9.71\%\\
18 - 23 & 24 & 36 &30.203 &2.070 &197 & 7.33\%\\
23 - 28 & 26 & 37 &32.323 &1.823 &269 & 10.01\%\\
28 - 33 & 30 & 37 &33.305 &1.470 &456 & 16.97\%\\
33 - 38 & 30 & 39 &33.781 &1.464 &410 & 15.26\%\\
38 - 43 & 31 & 39 &34.102 &1.327 &333 & 12.39\%\\
43 - 48 & 32 & 40 &34.549 &1.550 &184 & 6.85\%\\
48 - 53 & 32 & 40 &34.461 &1.562 &76 & 2.83\%\\
53 - 58 & 31 & 36 &34.381 &1.161 &21 & 0.78\%\\
58 - 61 & 33 & 36 &34.8 &1.304 &5& 0.19\%\\
  \hline
Total & 12 & 40 & 30.77 & 4.87 & 2687 & 100\%\\
  \hline
\end{tabular}
\end{center}
\end{table}

\begin{figure}[t!]
\centering
\includegraphics[width=5cm,height=5cm]{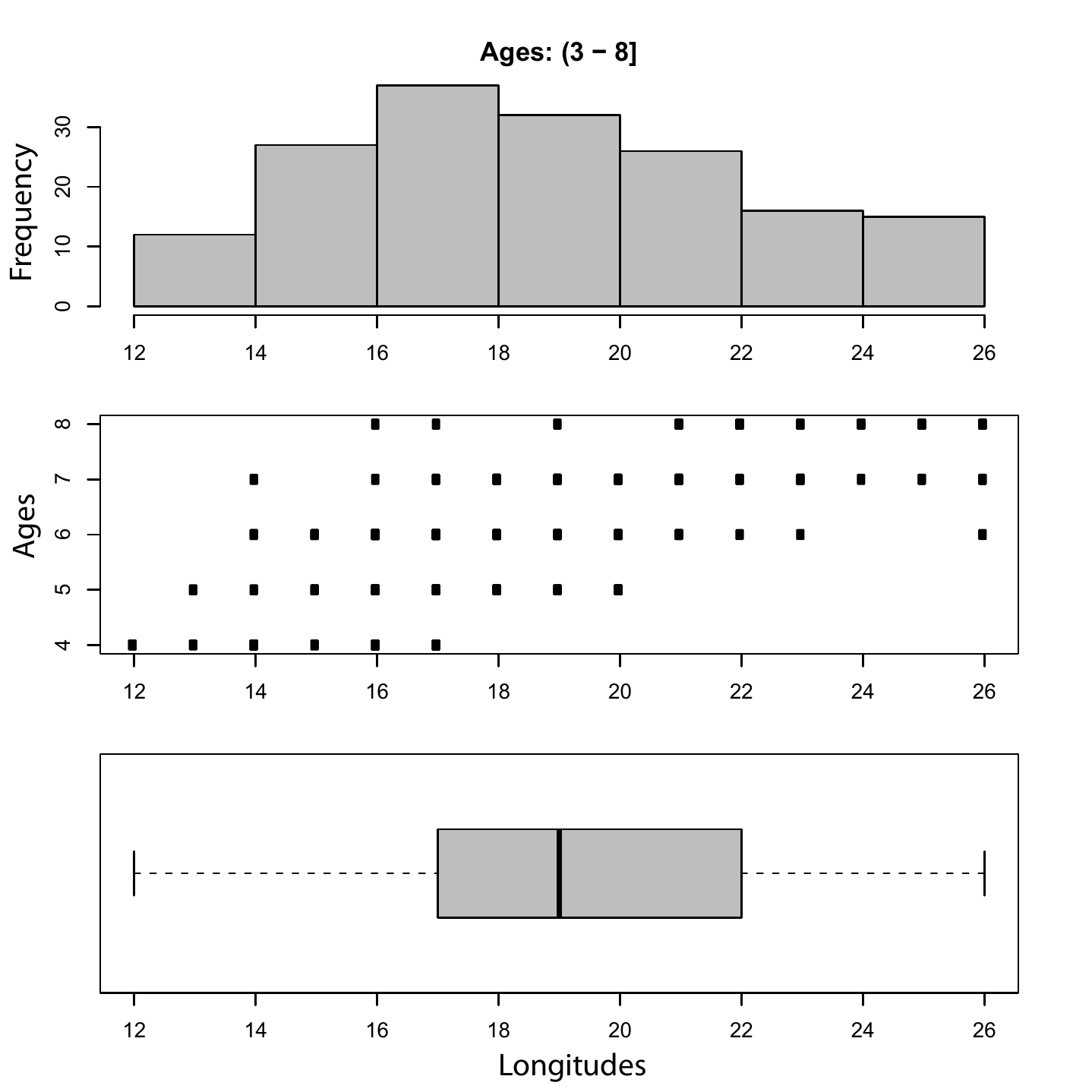}\hskip 7.5mm
\includegraphics[width=5cm,height=5cm]{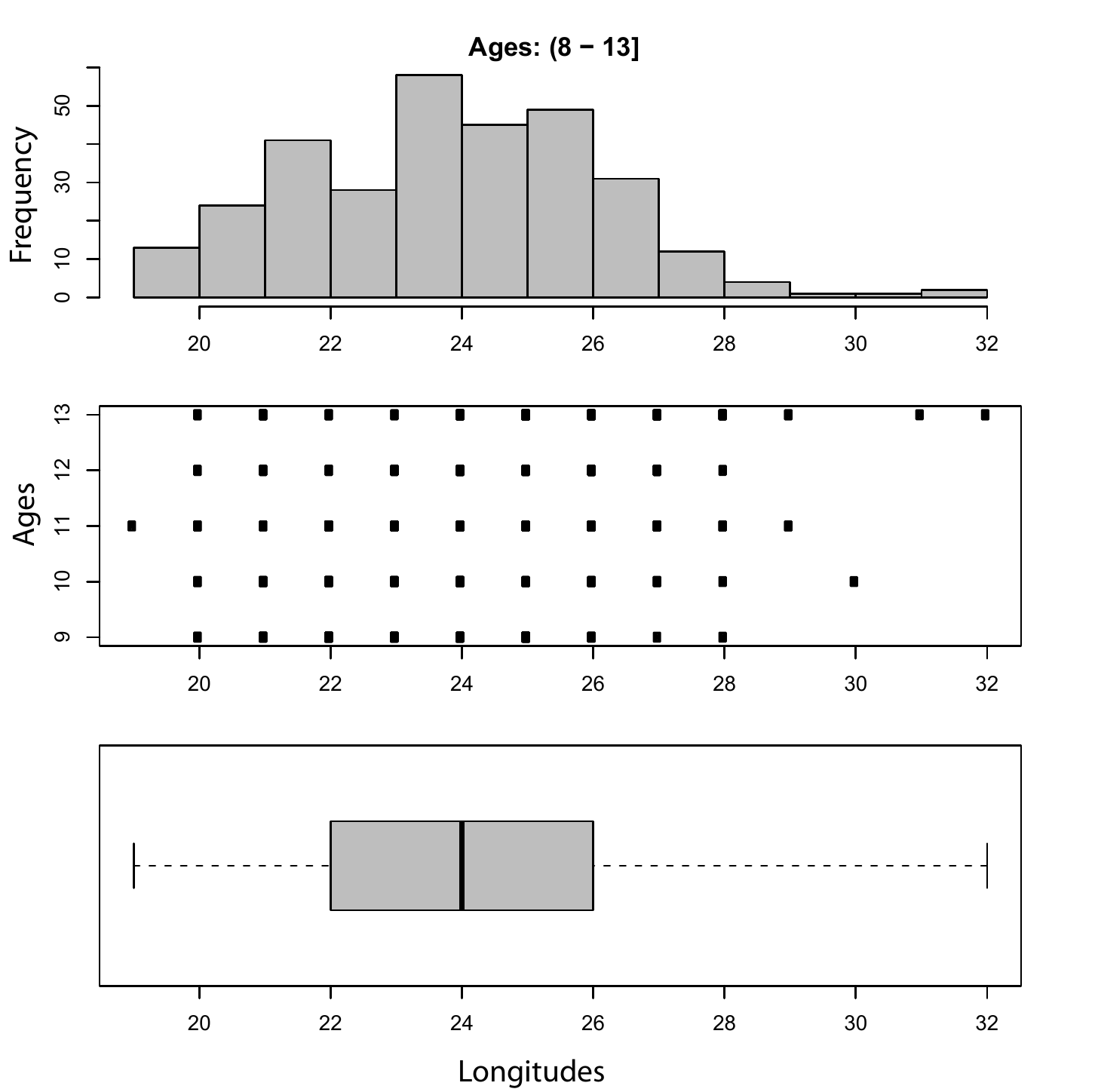}\\
\vskip 7.5mm
\includegraphics[width=5cm,height=5cm]{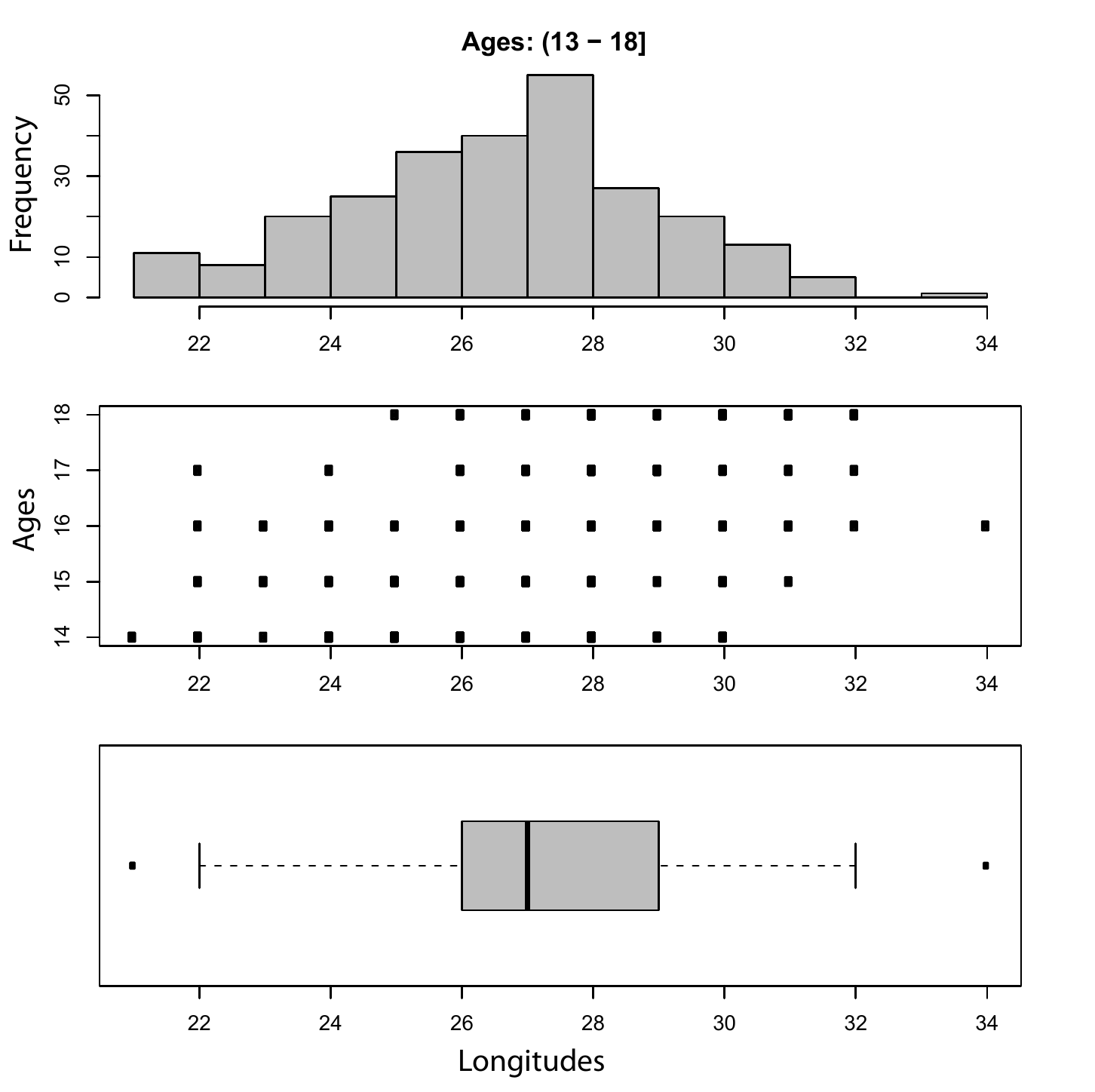}\hskip 7.5mm
\includegraphics[width=5cm,height=5cm]{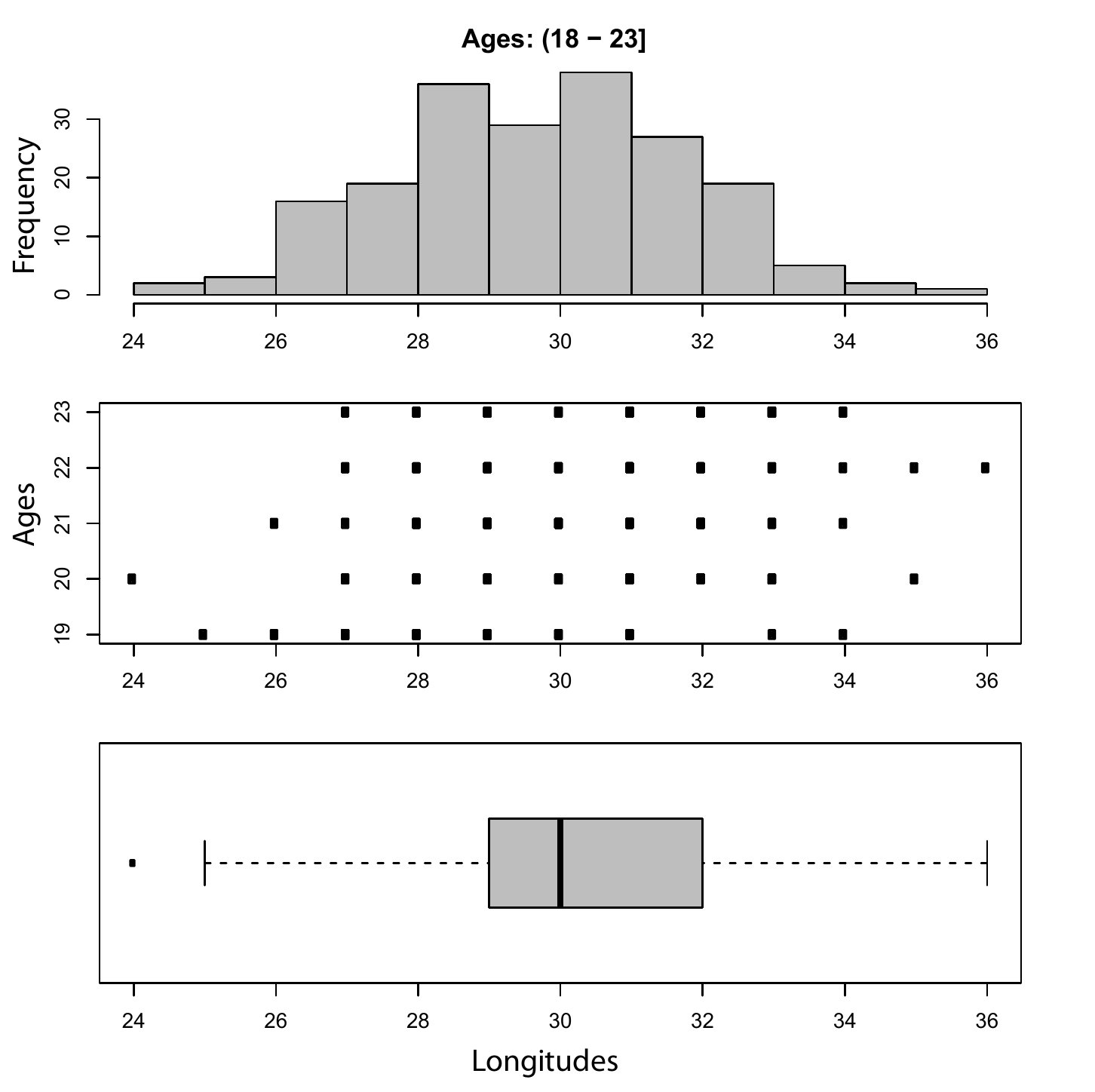}\\
\vskip 7.5 mm
\includegraphics[width=5cm,height=5cm]{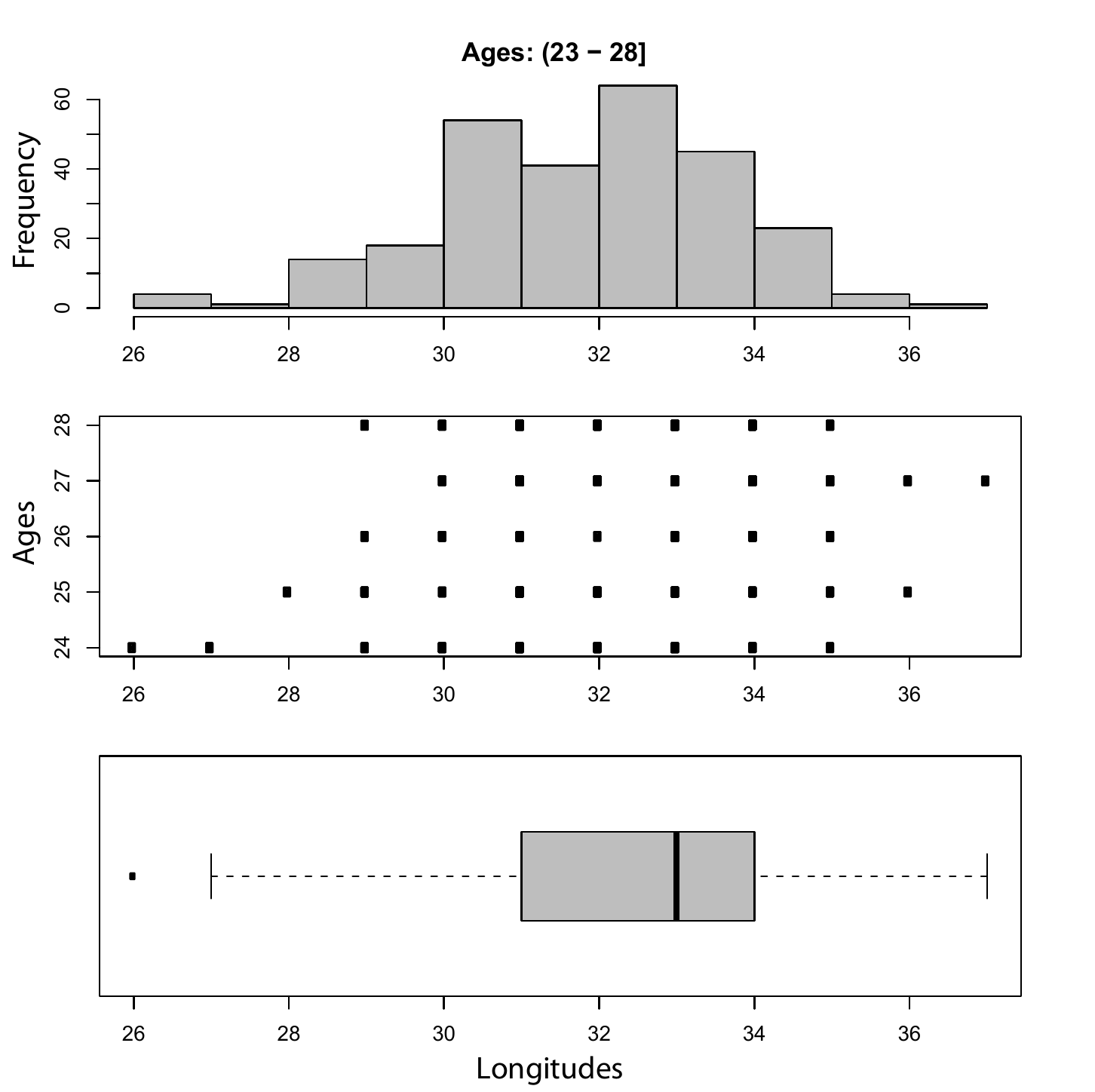}\hskip 7.5mm
\includegraphics[width=5cm,height=5cm]{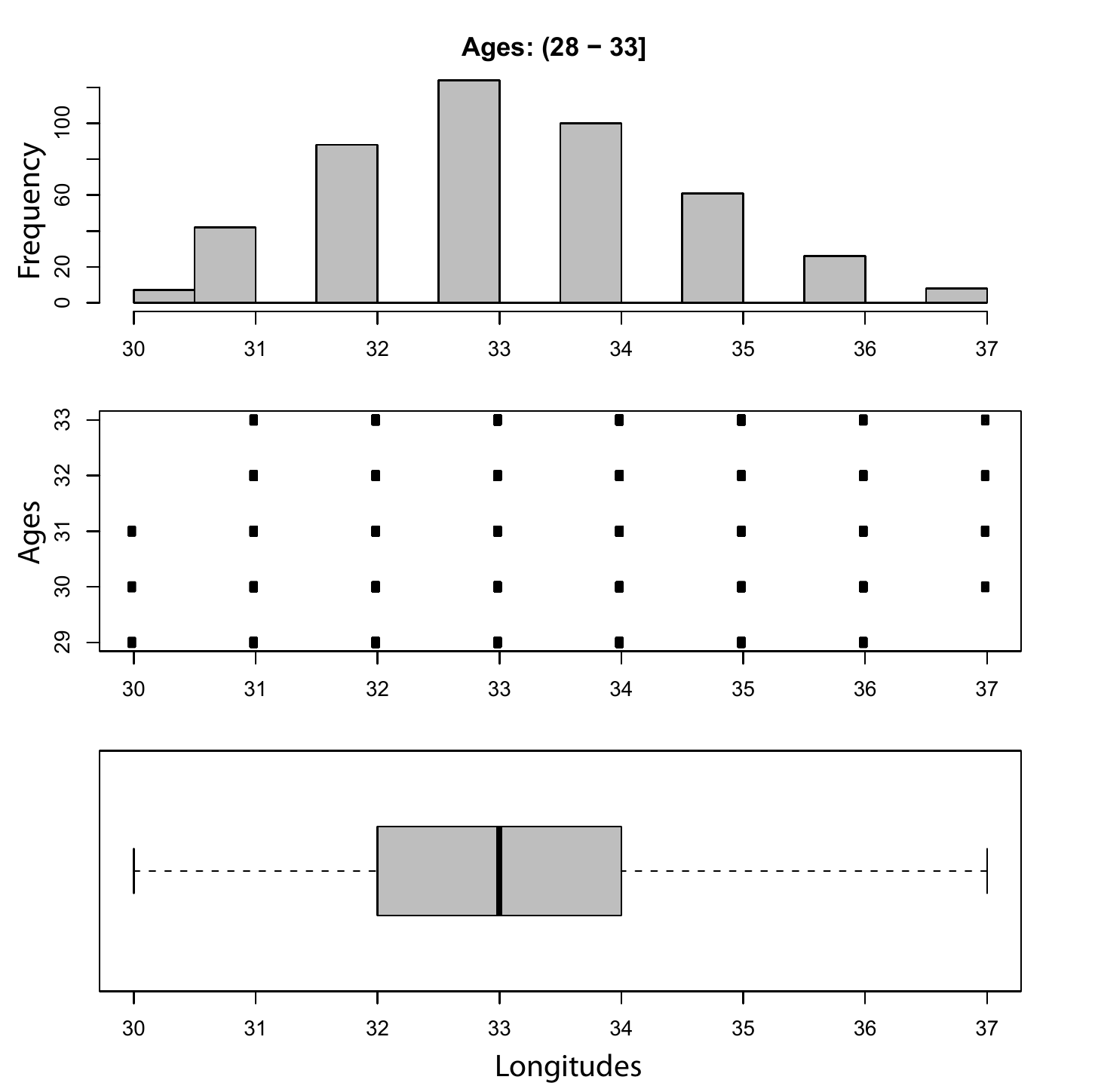}
\vskip 7.5 mm
    \caption{Histograms, Scatter-plots and Box-plots of lengths such age categories [3-8], (8-13], (13-18], (18-23], (23-28] and (28-33] for cardinalfish.}\label{H1}
\end{figure}


\begin{figure}[t!]
\centering
\includegraphics[width=5cm,height=5cm]{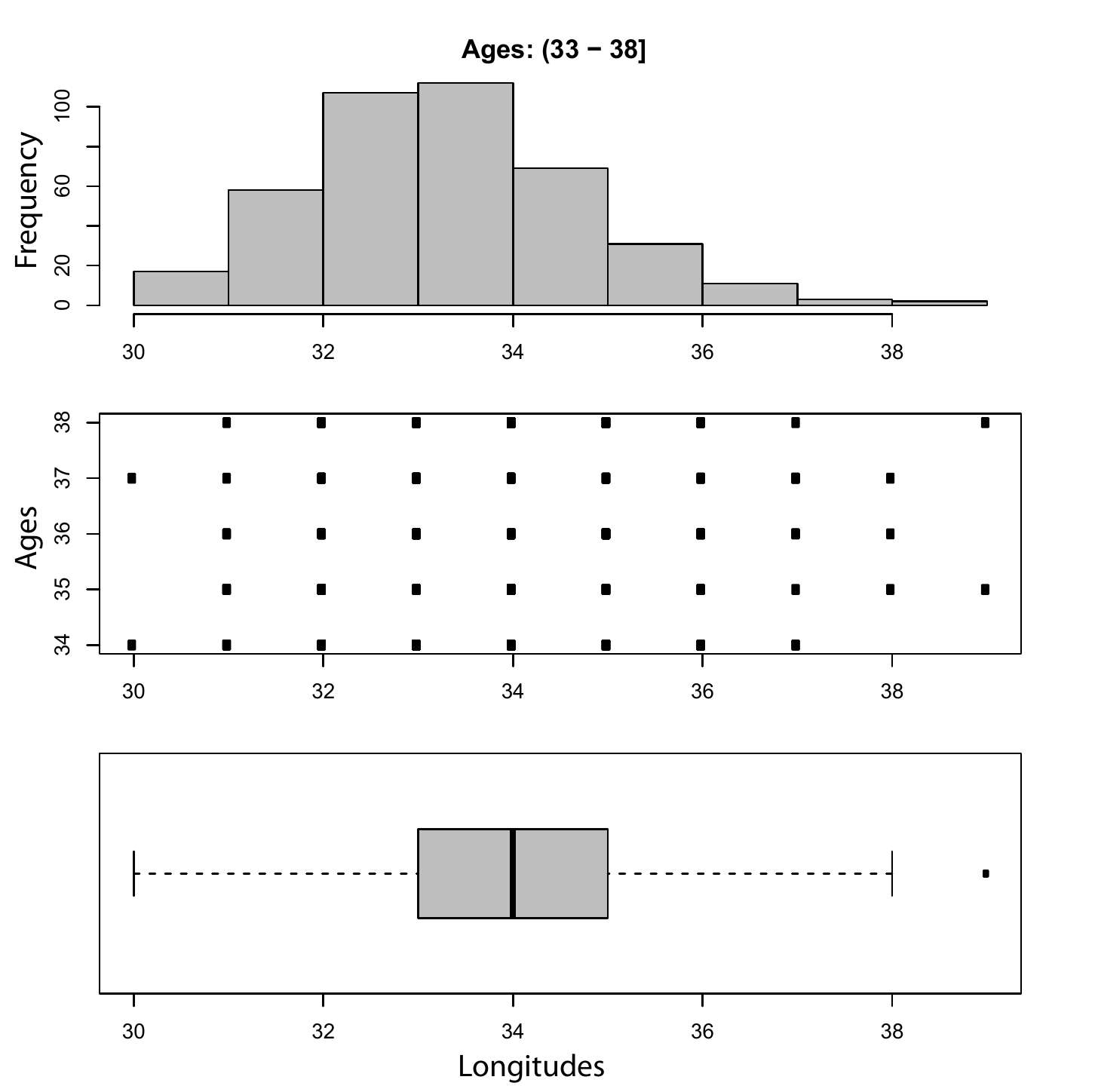}\hskip 7.5mm
\includegraphics[width=5cm,height=5cm]{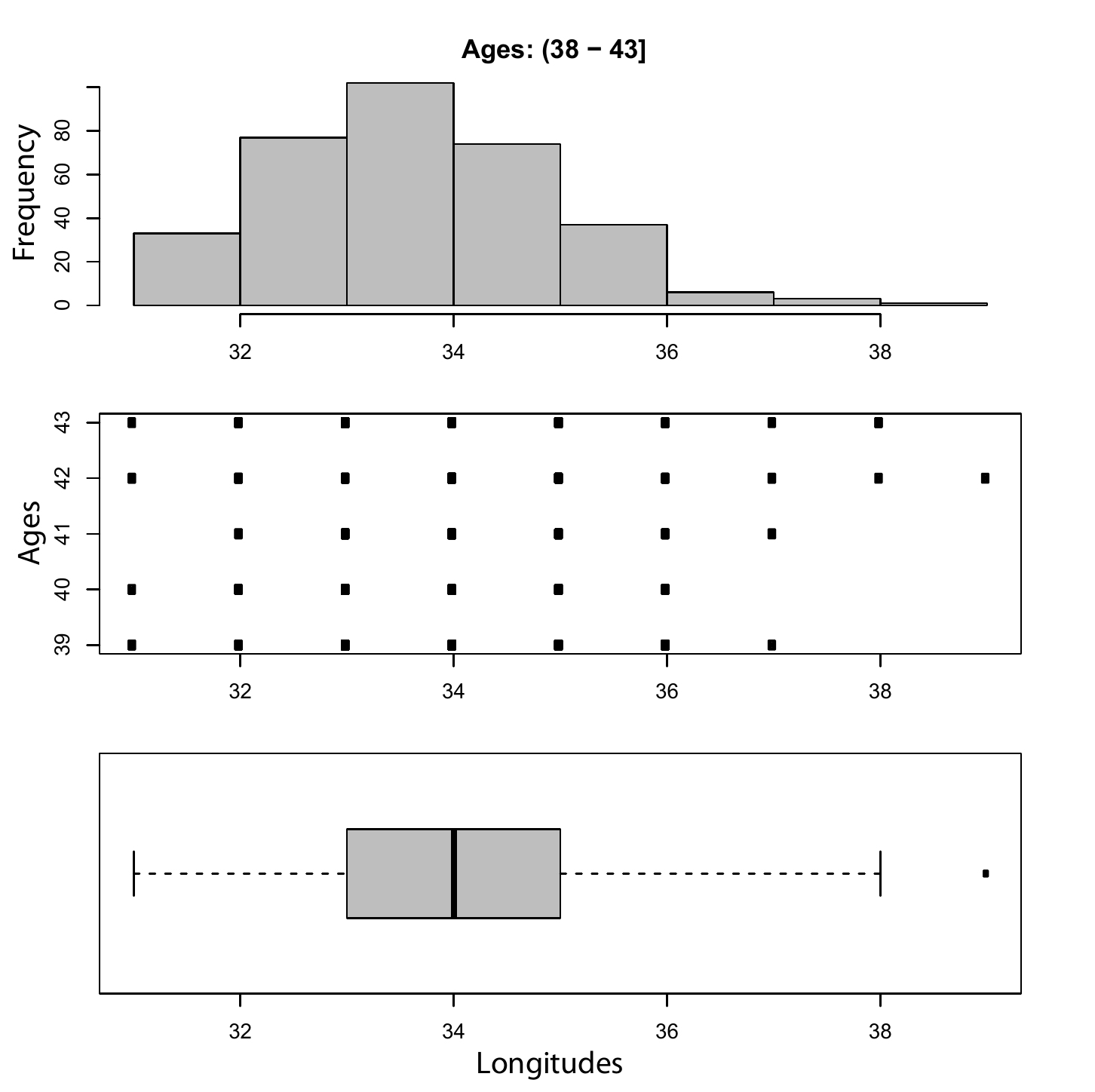}\\
\vskip 7.5 mm
\includegraphics[width=5cm,height=5cm]{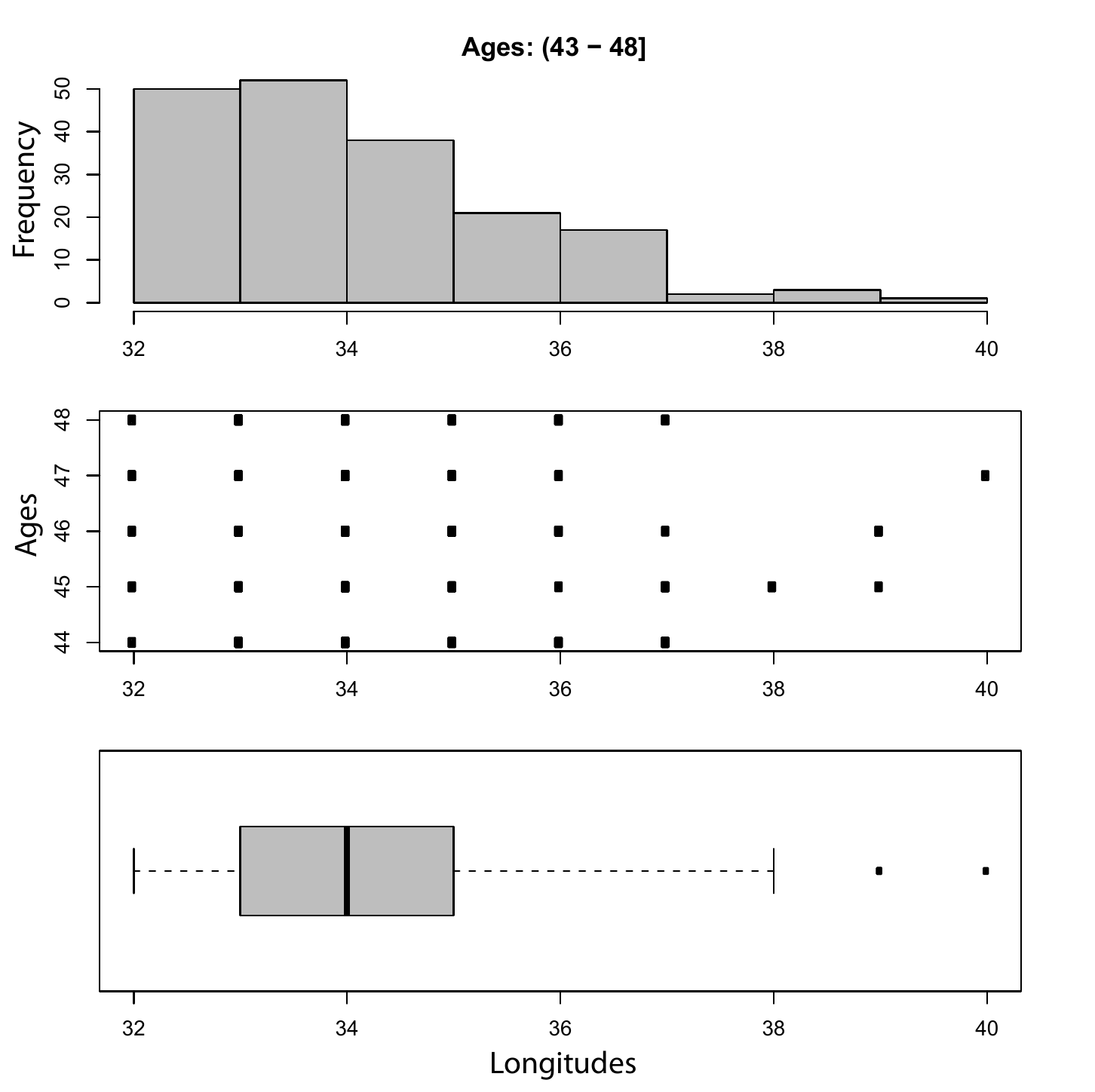}\hskip 7.5mm
\includegraphics[width=5cm,height=5cm]{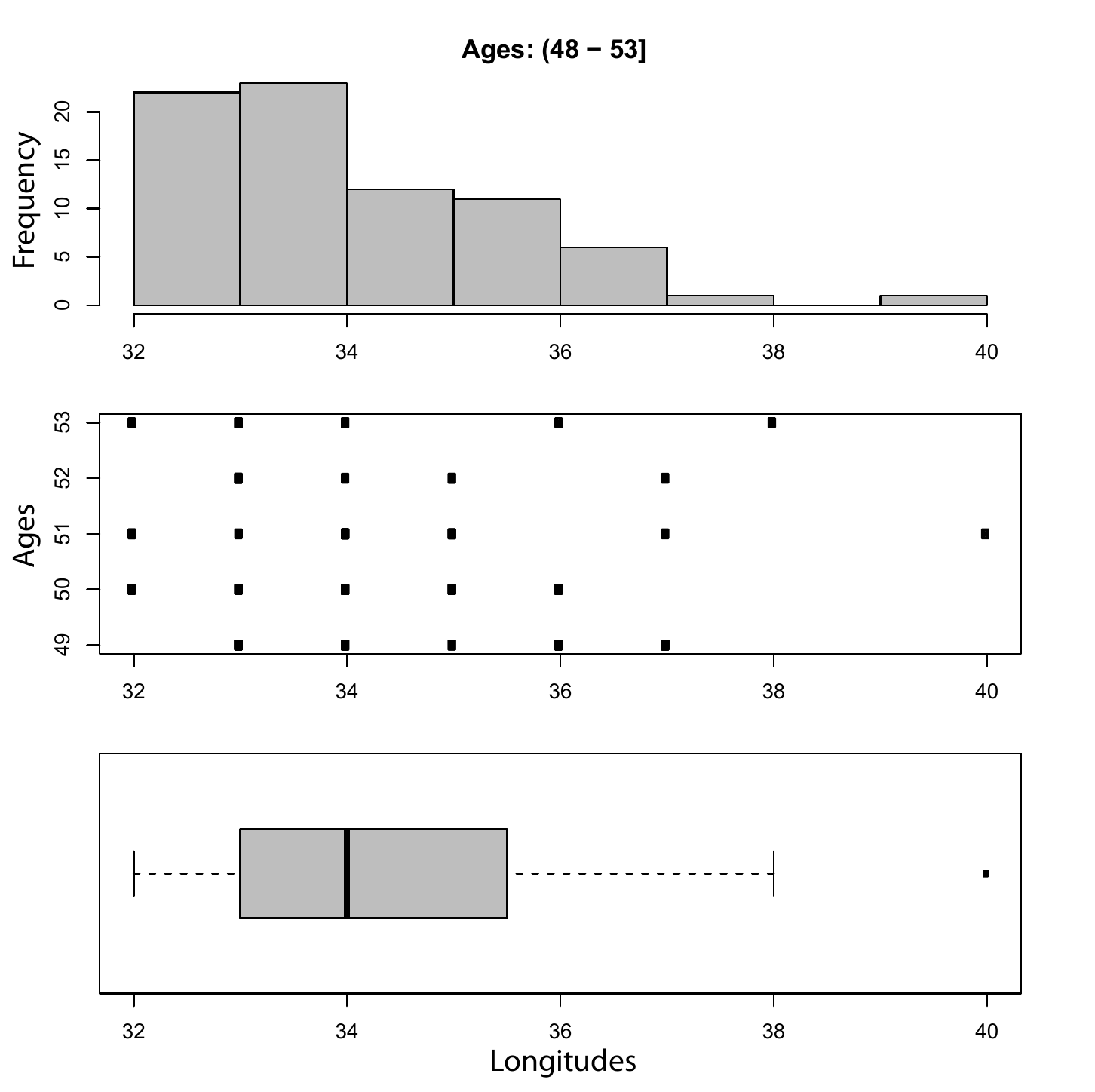}\\
\vskip 7.5 mm
\includegraphics[width=5cm,height=5cm]{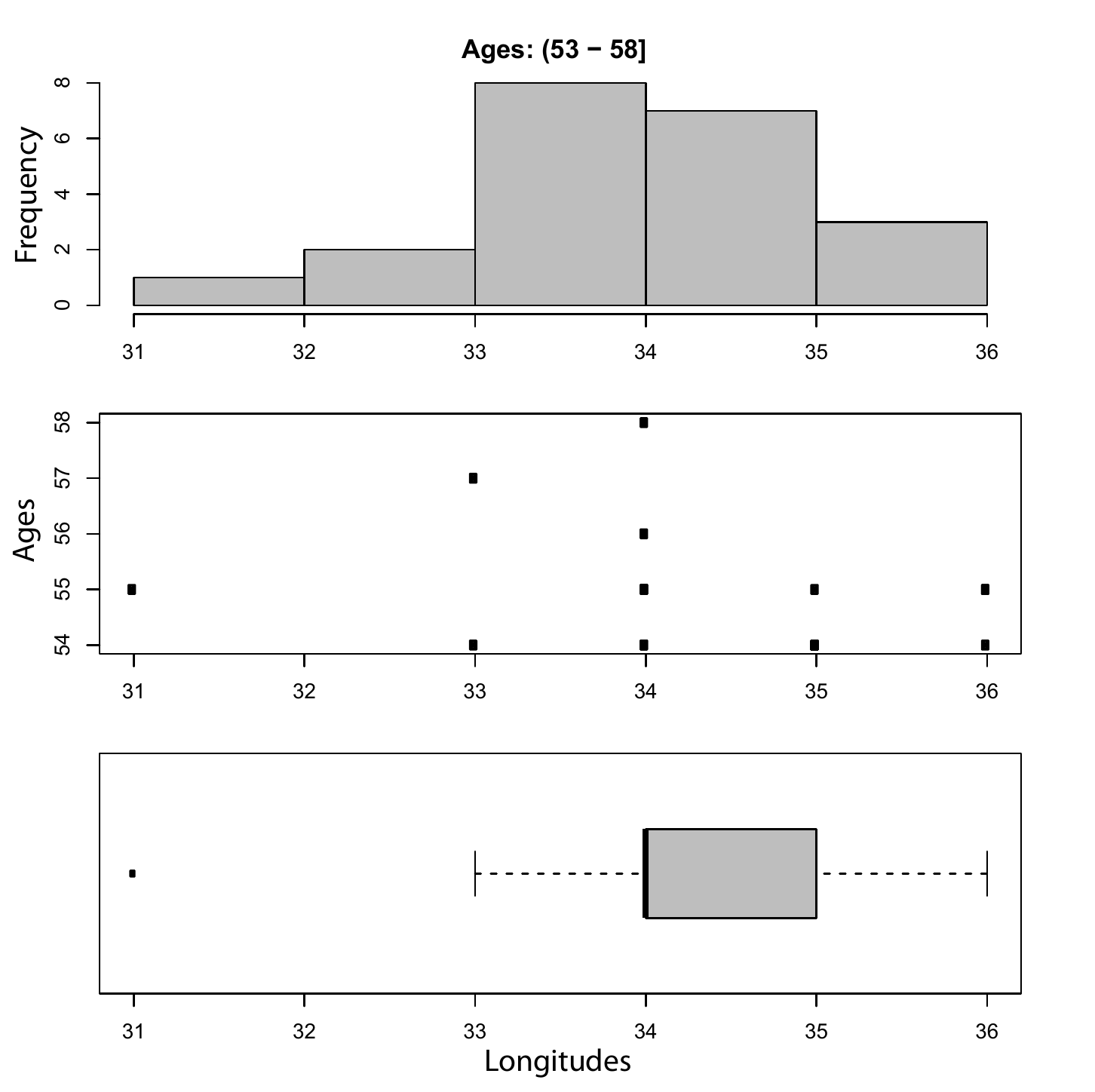}\hskip 7.5mm
\includegraphics[width=5cm,height=5cm]{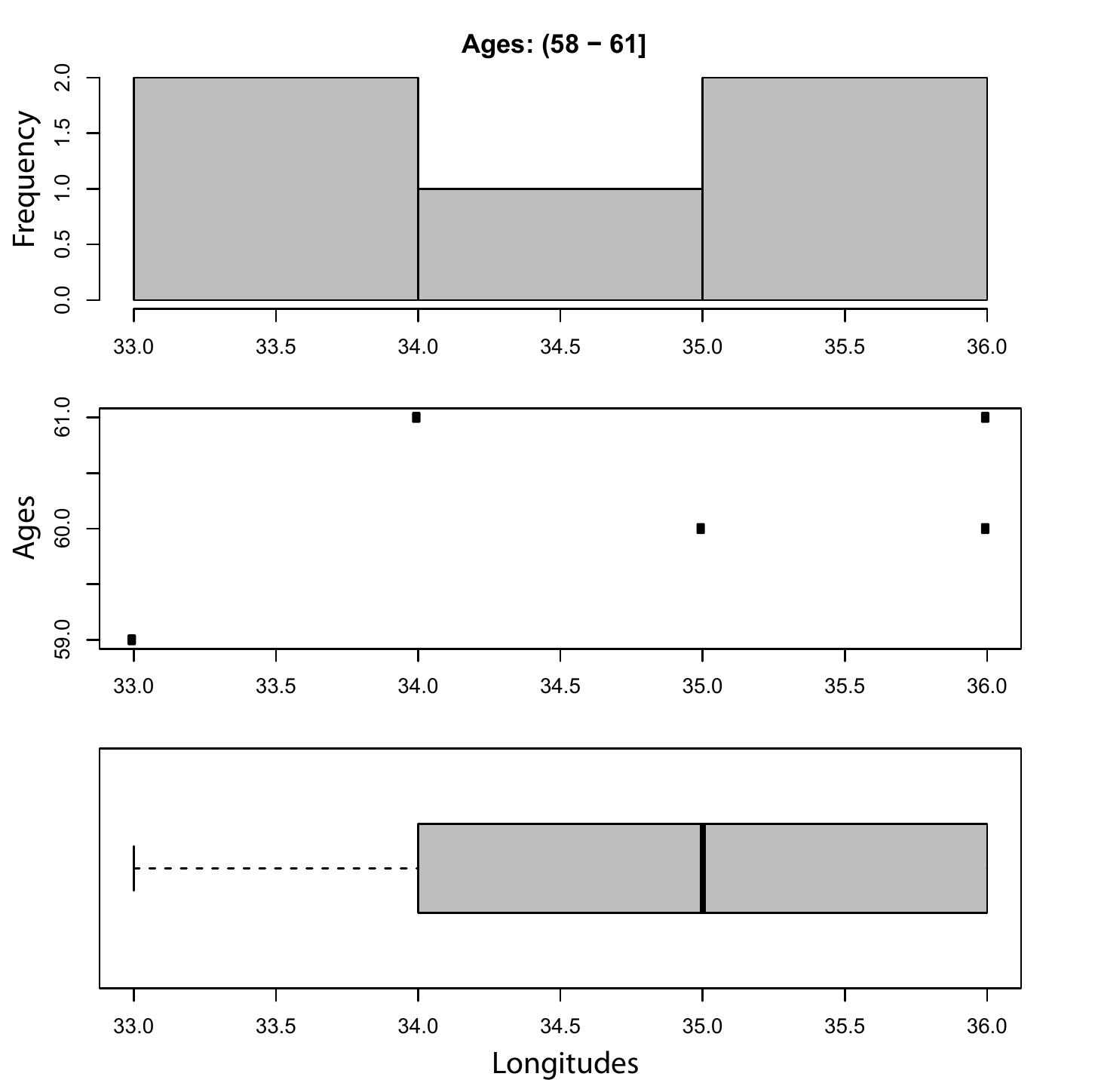}
\vskip 7.5 mm
    \caption{Histograms, Scatter-plots and Box-plots of lengths such age categories (33-38], (38-43], (43-48], (48-53], (53-58] and (58-61] for cardinalfish.}\label{H2}
\end{figure}


\begin{table}[htb]
\caption{T model fits for $\nu=2,5,...,45$ considering the full data and the data without influential observations (IO).}\label{CG2}

\begin{center}
\begin{small}
\begin{tabular}{cccccccc}
  \hline
 & & 2 & 5 & 15 &  25 &  35 &  45  \\
  \hline
\multirow{9}{*}{\rotatebox{90}{Full Data}} &    $L_{\infty}$ & 34.897 & 34.993 & 35.075 & 35.099 & 35.111 & 35.118  \\
&$\mbox{SE}(L_{\infty})$ & 0.060 & 0.073  & 0.082  & 0.084  & 0.086  & 0.086  \\
  &$K$ & 0.087 & 0.085  & 0.084 & 0.084 & 0.083  & 0.083  \\
  &$\mbox{SE}(K)$ & 0.001 & 0.002  & 0.002  & 0.002  & 0.002  & 0.002  \\
  &$t_0$ & -2.727 & -2.858  & -2.968  & -3.002  & -3.019  & -3.029  \\
 &$\mbox{SE}(t_0)$ & 0.201 & 0.240  & 0.269  & 0.277  & 0.280  & 0.282  \\
  &$N$ & 2687& 2687  & 2687 & 2687  & 2687 & 2687  \\
  &AIC & 10951.699 & 10663.449  & 10593.598  & 10588.949  & {\bf 10588.359}  & 10588.445  \\
  &$\ell(\btheta)$ & -5470.849 & -5326.725  & -5291.799  & -5289.475  & {\bf -5289.180}  & -5289.223  \\
  \hline
\multirow{10}{*}{\rotatebox{90}{Without IO}} &  $L_{\infty}$ & 34.889 & 34.976 & 35.046 & 35.065 & 35.074 & 35.079 \\
&$\mbox{SE}(L_{\infty})$ & 0.060 & 0.072 & 0.080 & 0.083 & 0.084 & 0.084 \\
  &$K$ & 0.087 & 0.086 & 0.085 & 0.084 & 0.084 & 0.084 \\
  &$\mbox{SE}(K)$ & 0.001 & 0.002 & 0.002 & 0.002 & 0.002 & 0.002 \\
  &$t_0$ & -2.692 & -2.781 & -2.848 & -2.864 & -2.872 & -2.876 \\
  &$\mbox{SE}(t_0)$ & 0.199 & 0.235 & 0.263 & 0.270 & 0.273 & 0.275 \\
  &$N$ & 2679 & 2678 & 2678 & 2678 & 2678 & 2678 \\
  &$\ell(\btheta)$ & -5434.569 & -5283.160 & -5244.821 & -5241.333 & {\bf -5240.446} & -5240.130 \\
  \hline
\multirow{3}{*}{\rotatebox{90}{RC}}    &$L_{\infty}$ & 0.023\% &  0.05\%  &  0.082\%  &  0.096\%  &  0.11\%  &  0.11\%  \\
&$K$ & 0\% &  1.17\%  &  1.19\%  &  0\%  &  1.21\%  &  1.21\%  \\
  &$t_0$ & 1.28\% &   2.69\%  &  4.04\%  &  4.59\%  &  4.87\%  &  5.05\%  \\
  \hline
\end{tabular}
\end{small}
\end{center}
\end{table}

\begin{table}[htb]
\caption{ST model fits for $\nu=2,5,...,45$ considering the full data and the data without influential observations (IO).}\label{CG3}

\begin{center}
\begin{small}
\begin{tabular}{cccccccc}
  \hline
 & & 2 & 5 & 15  & 25  & 35 & 45 \\
  \hline
\multirow{9}{*}{\rotatebox{90}{Full Data}} & $L_{\infty}$ & 35.025 & 35.038  & 35.098  & 35.119  & 35.128  & 35.134  \\
  &$\mbox{SE}(L_{\infty})$ & 0.059 & 0.069  & 0.071  & 0.07  & 0.069  & 0.068  \\
  &$K$ & 0.086 & 0.085  & 0.084  & 0.083  & 0.083  & 0.083  \\
    &$\mbox{SE}(K)$ & 0.001 & 0.002  & 0.002  & 0.002  & 0.002  & 0.001  \\
  &$t_0$ & -2.896 & -2.937  & -3.020  & -3.049  & -3.063  & -3.071  \\
  &$\mbox{SE}(t_0)$ & 0.199 & 0.230 & 0.236  & 0.231  & 0.227  & 0.225  \\
  &$N$ & 2687 & 2687  & 2687 & 2687 & 2687 & 2687 \\
  &AIC & 10947.539 & 10659.820  & 10589.693  & 10584.585  & 10583.607  & 10583.392  \\
  &$\ell(\btheta)$ & -5467.769 & -5323.910 &  -5288.846  & -5286.293  & -5285.803  & -5285.696  \\
  \hline
  \multirow{10}{*}{\rotatebox{90}{Without IO}} & $L_{\infty}$ & 35.033 & 35.050 &  35.094  & 35.093  & 35.090  & 35.095  \\
  &$\mbox{SE}(L_{\infty})$ & 0.059 & 0.069  & 0.071  & 0.071  & 0.071  & 0.07  \\
  &$K$ & 0.086 & 0.084  & 0.084  & 0.084& 0.084 & 0.084  \\
  &$\mbox{SE}(K)$ & 0.001 & 0.002  & 0.002  & 0.002  & 0.002  & 0.002  \\
  &$t_0$ & -2.922 & -3.016 & -2.907  & -2.889  & -2.876  & -2.882  \\
  &$\mbox{SE}(t_0)$ & 0.202 & 0.231  & 0.233  & 0.232  & 0.233  & 0.231  \\
  &$N$ & 2677 & 2684  & 2680 & 2678  & 2677  & 2677  \\
  &$\ell(\btheta)$ & -5439.118 & -5314.756  & -5258.536  & -5244.444  & -5236.042  & -5235.694  \\
    \hline
  \multirow{3}{*}{\rotatebox{90}{RC}}&$L_{\infty}$ & 0.02\% &  0.03\%  & 0.01\%  & 0.07\%  & 0.11\%  & 0.11\%  \\
  &$K$ & 0\% & 1.19\%  &  0\% &  1.19\%  &  1.19\%  &  1.19\%  \\
  &$t_0$ & 0.89\% &  2.62\% & 3.89\% & 5.54\%  & 6.50\%  & 6.56\%  \\
  \hline
\end{tabular}
\end{small}
\end{center}
\end{table}

\begin{figure}[t!]
  \centering
  \includegraphics[scale=0.5]{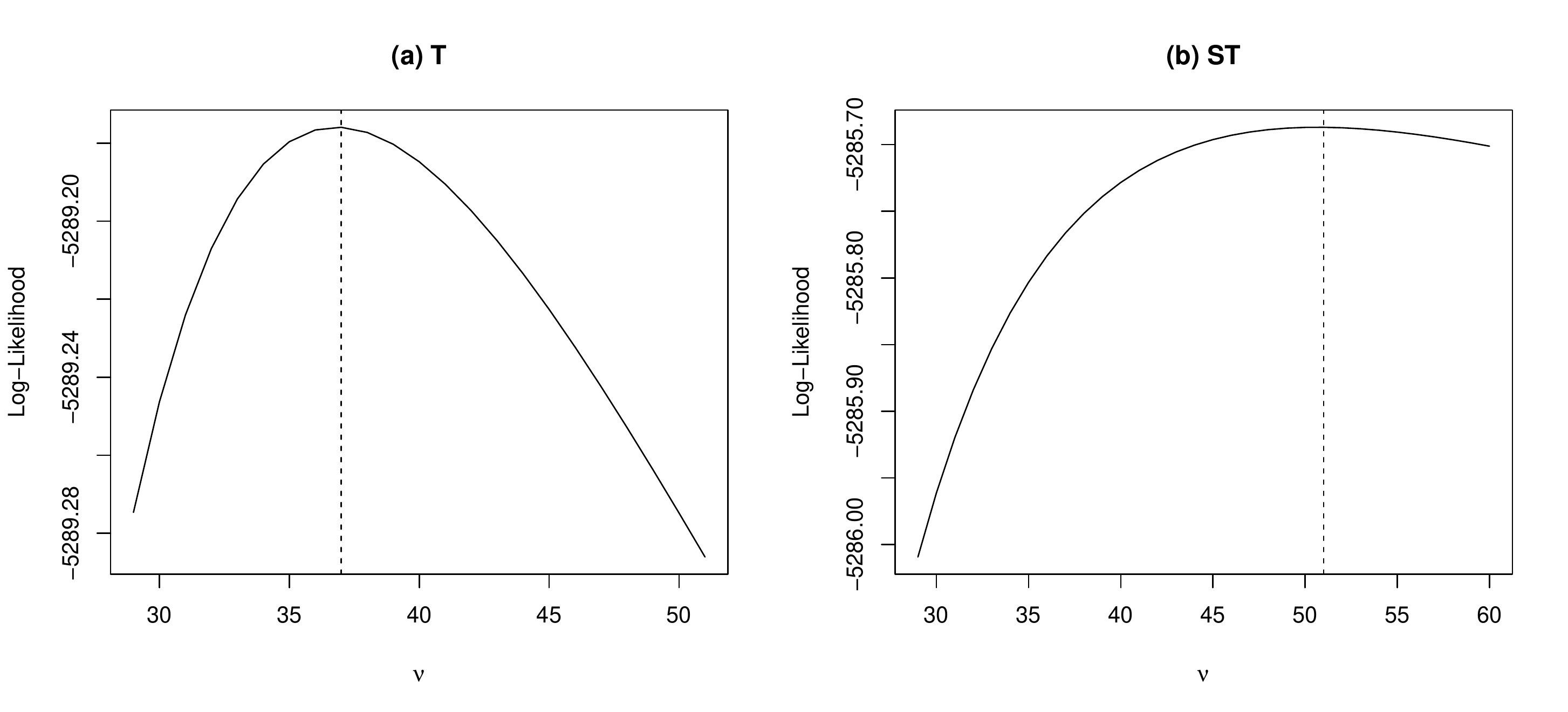}
  \caption{(a) Profiles Log-Likelihood for T fit model. The dashed line corresponds to maximum log-likelihood at $\nu=37$.
  (b) Profiles Log-Likelihood for ST fit model. The dashed line corresponds to maximum log-likelihood at $\nu=51$.}\label{LogL}
\end{figure}


\begin{figure}[t!]
\centering
\includegraphics[width=6cm,height=6cm]{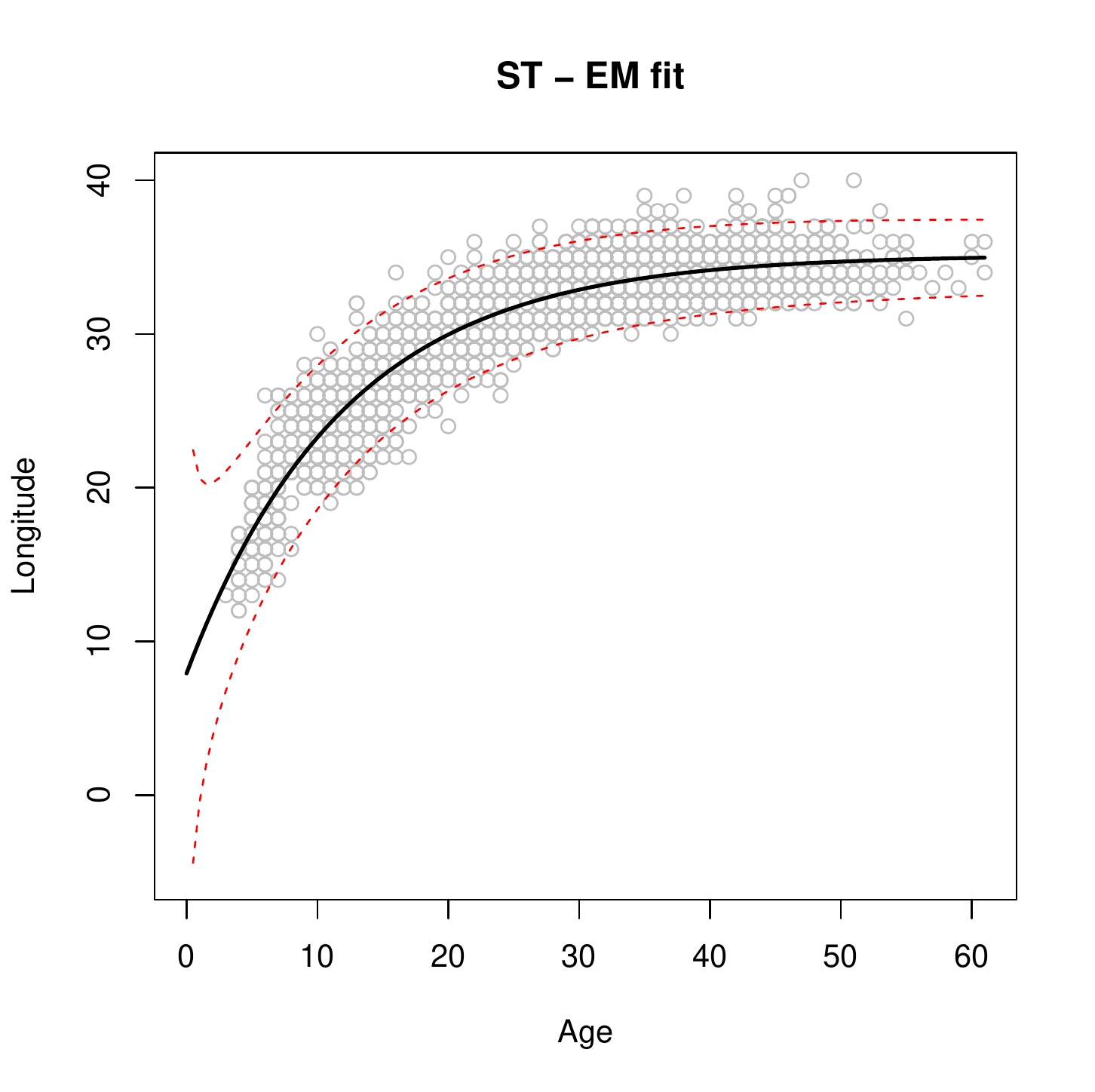}
\includegraphics[width=6cm,height=6cm]{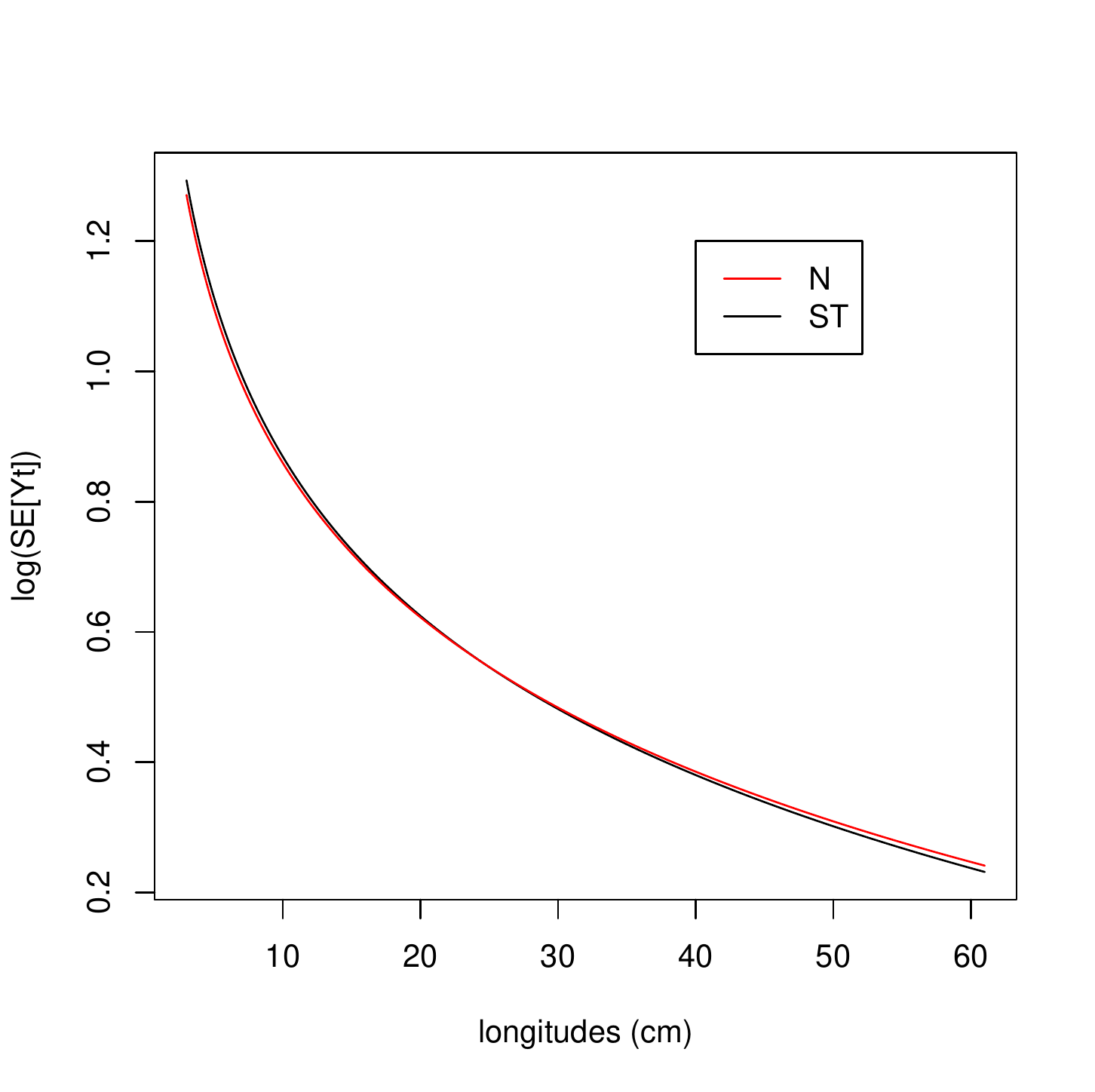}
    \caption{Left: Cardinalfish observations. The thick line corresponds to the VB fit using ST distribution for $\nu=51$ and
    the dashed corresponds to confidence intervals at 5\% significance level. Right: $\log\mbox{ }\sqrt{Var[y_t]}$ values for N and ST distributions.}\label{K22}
\end{figure}


\begin{table}[htb]
\caption{Summary of T (with $\nu=37$) and ST (with $\nu=51$) model fits considering the full data and the data without influential observations (IO).}\label{CG4}
\begin{center}
\begin{tabular}{ccccc}
  \hline
  & \multicolumn{2}{c}{Complete Data} & \multicolumn{2}{c}{Without IO} \\
  \hline
  Parameter & T & ST & T & ST \\
  \hline
  $L_{\infty}$ & 35.113 & 35.137 & 35.095 & 35.097\\
  & (0.086) & (0.068) & (0.085) & (0.07)\\
  $K$ & 0.083 & 0.083 &  0.084 & 0.084\\
  & (0.002) & (0.001) & (0.002) & (0.001)\\
  $t_0$ & -3.021 & -3.075 & -2.943 & -2.884\\
  & (0.281) & (0.224) & (0.277) & (0.23)\\
  $\rho$ & -0.690 & -0.705 & -0.688 & -0.703\\
  $\sigma^2$ & 25.961 & 38.087 & 25.359 & 34.868\\
  $\lambda$ & - & 0.873 & - & 0.755\\
  $\nu$ & 37 & 51 & 37 & 51 \\
  $\kappa_1$ & 1.021 & 1.015 & 1.021 & 1.015 \\
  $\kappa_2$ & 1.057 & 1.041 & 1.057 & 1.041 \\
  \hline
  $\ell(\btheta)$ & -5289.176 & -5285.687 & -5248.69 & -5235.583\\
  $n$ & 2687 & 2687 & 2680 & 2677\\
  AIC & 10588.35 & 10583.37 & 10507.38 & 10483.17\\
  \hline
\end{tabular}
\end{center}
\end{table}

\begin{figure}[htb]
\centering
\includegraphics[width=6cm,height=6cm]{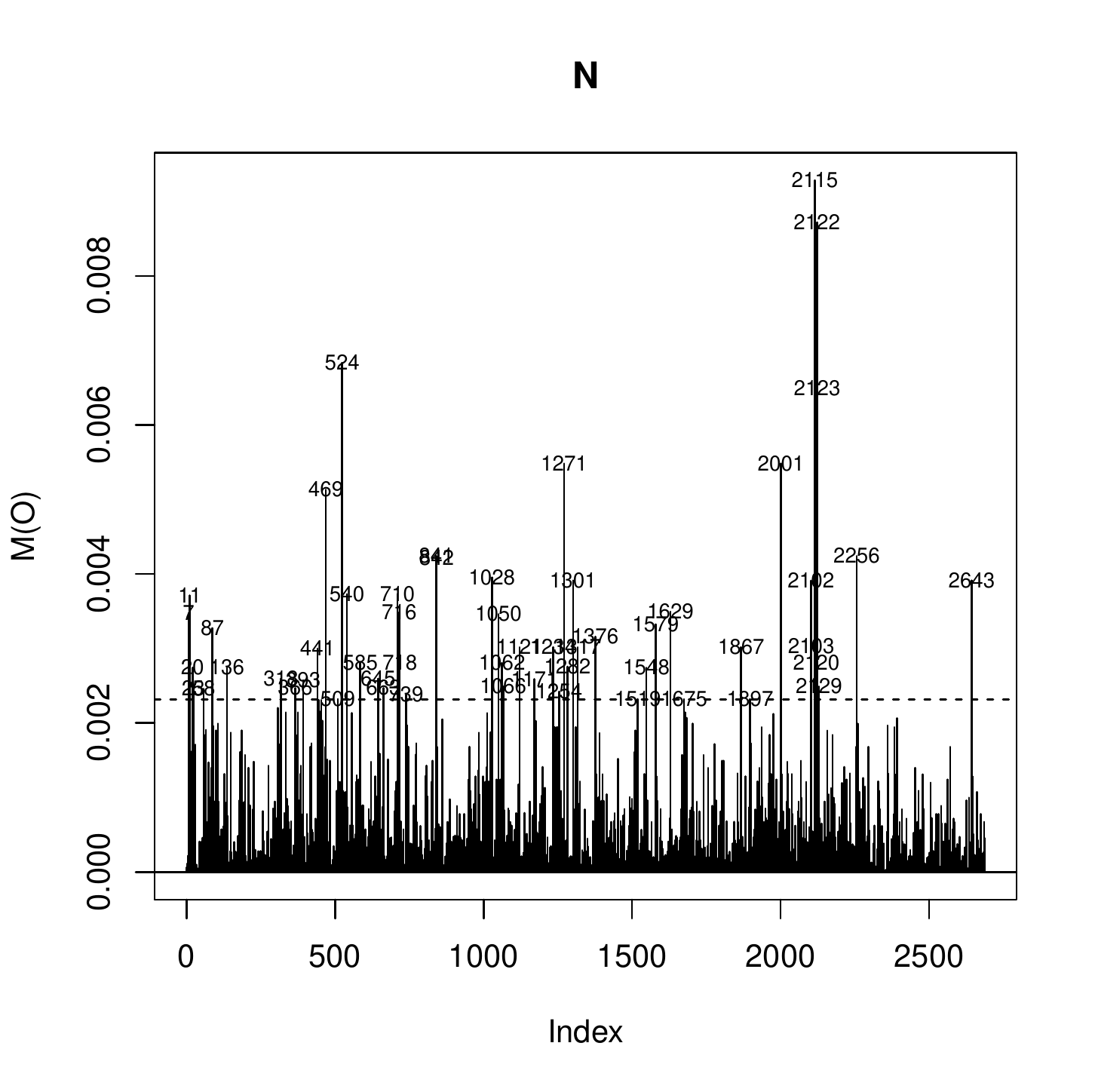}
\includegraphics[width=6cm,height=6cm]{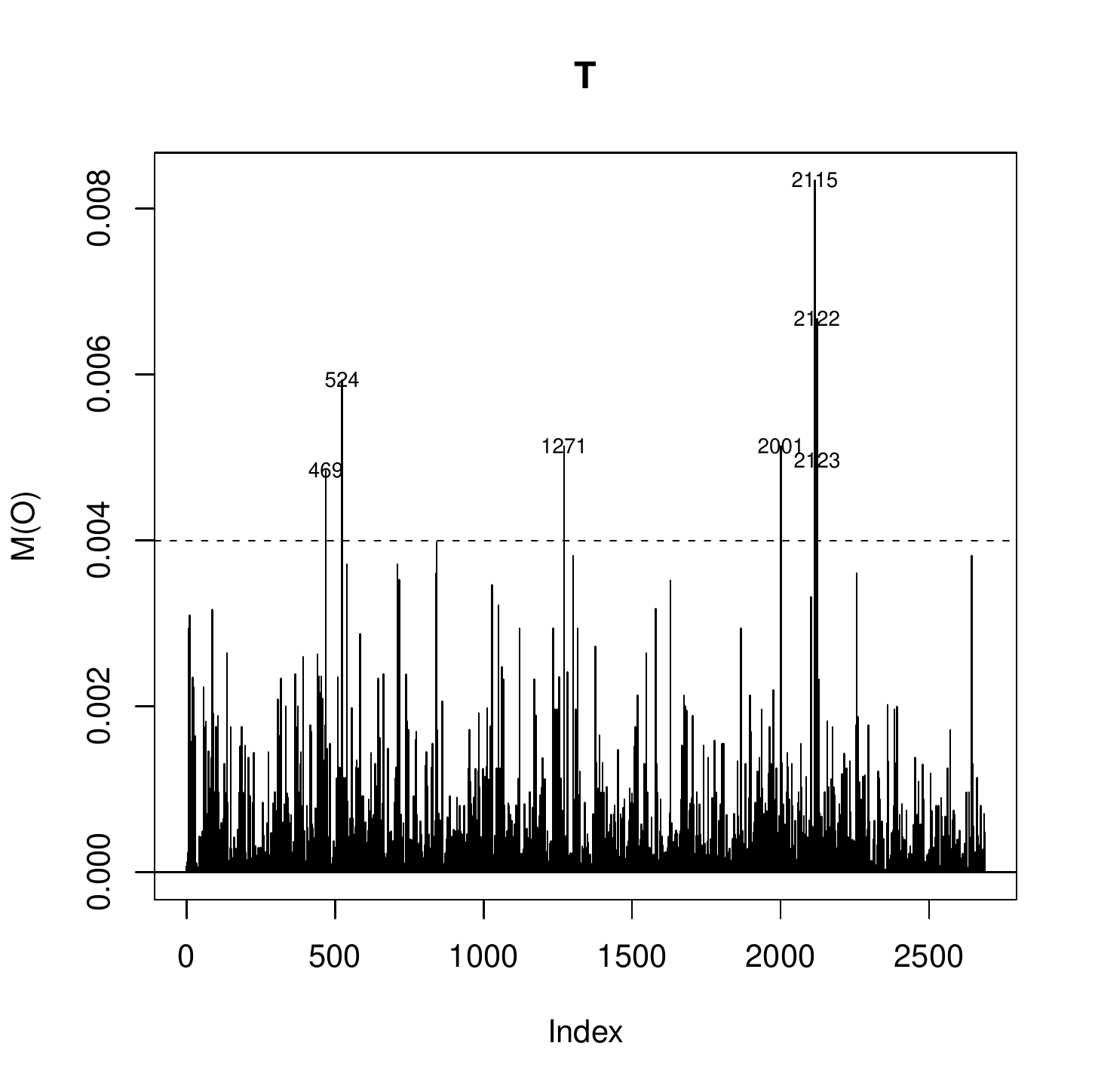}\\
\includegraphics[width=6cm,height=6cm]{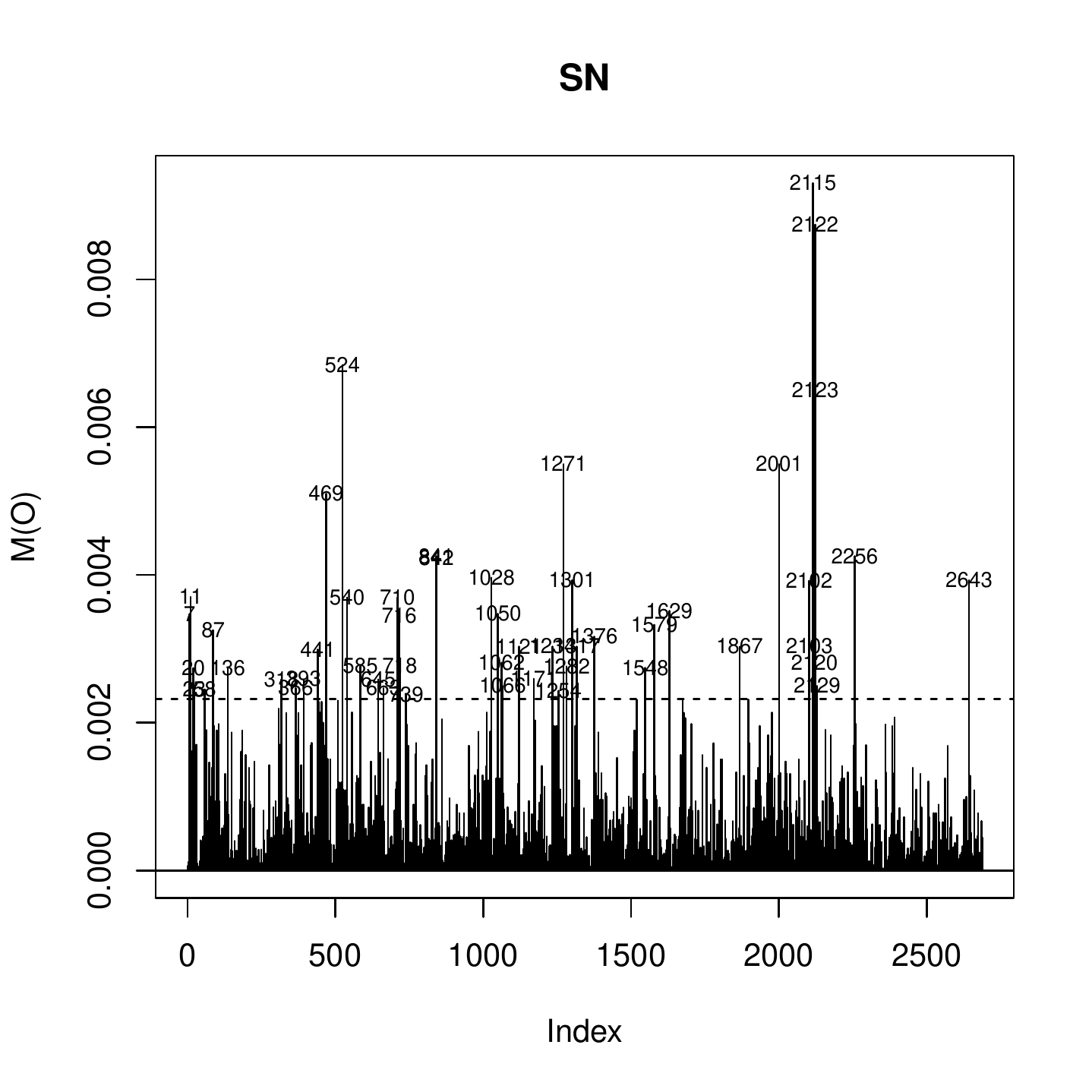}
\includegraphics[width=6cm,height=6cm]{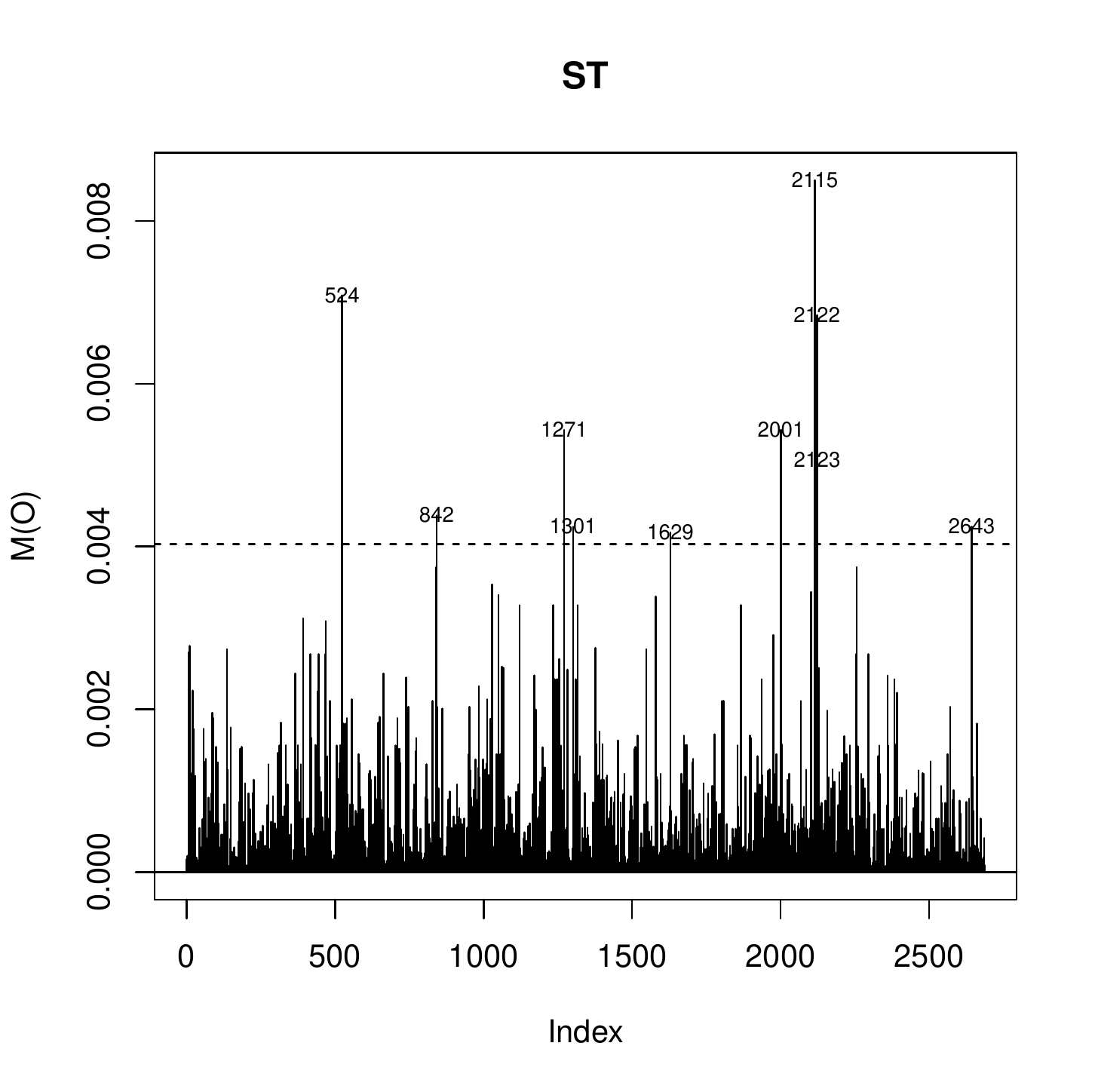}
\caption{Plots of Influence Analysis of Cook for the four distributions, the cases T and ST assume $\nu=37$ and $\nu=51$
degrees of freedom, respectively. The dotted line corresponds to the decision level $c_{\bd}$, the {\tt Index} axis
corresponds to the observation index and {\tt M(0)} to the curve value $\overline{M}_{0}(\btheta)$.
The enumerated lines correspond to the observation indexes where $\overline{M}_{0}(\btheta)\geq c_{\bd}$
is accomplished.}\label{K3}
\end{figure}
%

\begin{table}[htb]
\caption{Estimations ($\widehat{\beta}$) and standard errors (SE) for full data and for the VB growth curve parameters
$\beta^\top=(L_{\infty}, K, t_0)$ associated to normal (N), $t$-Student (T), skew-normal (SN) and skew-$t$ (ST) fits.
The T and ST cases assume $\nu=37$ and $\nu=51$ degrees of freedom, respectively. The notations $\widehat{\beta}_0$
and $\mbox{SE}_0$ represent the corresponding estimations and standard errors with the dataset without influential
observations.
}\label{CG1}
\begin{center}
\begin{tabular}{ccccccccc}
  \hline
  &  & $L_{\infty}$ & $K$ & $t_0$ & $n$ & $q$ & $\ell(\btheta)$ & AIC \\
  \hline
   & $\widehat{\beta}$ & 35.147 & 0.083 & -3.072 & 2687 & 5 & -5291.7 & 10591.4 \\
   & SE & 0.089 & 0.002 & 0.290 & - & - & - & - \\
  N & $\widehat{\beta}_0$ & 35.103 & 0.084 & -2.922 & 2633 & 5 &  &  \\
      & $\mbox{SE}_0$ & 0.085 & 0.002 & 0.269 & - & - & - & - \\
   & RC & 0.1\% & 1.2\% & 5\% & 2.01\% &  &  &  \\
   \hline
   & $\widehat{\beta}$ & 35.113 & 0.083 & -3.021 & 2687 & 5 & -5289.18 & 10588.35\\
   & SE & 0.086 & 0.002 & 0.281 & - & - & - & - \\
  T & $\widehat{\beta}_0$ & 35.095 &  0.084 & -2.943 & 2680 & 5 &  &  \\
         & $\mbox{SE}_0$ & 0.085 & 0.002 & 0.277 & - & - & - & - \\
   & RC & 0.05\% & 1.21\% & 2.58\% & 0.26\% &  &  &  \\
   \hline
   & $\widehat{\beta}$ & 35.147 & 0.083 & -3.071 & 2687 & 6 & -5290.7 & 10593.4 \\
   & SE & 0.089 & 0.002 & 0.290 & - & - & - & - \\
  SN & $\widehat{\beta}_0$ & 35.109 &  0.084 & -2.905 & 2637 & 6 &  &  \\
         & $\mbox{SE}_0$ & 0.085 & 0.002 & 0.271 & - & - & - & - \\
   & RC & 0.1\% & 1.2\% & 5.4\% & 1.86\% &  &  &  \\
   \hline
   & $\widehat{\beta}$ & 35.137 & 0.083 & -3.075 & 2687 & 6 & -5285.69 & 10583.37 \\
   & SE & 0.085 & 0.002 & 0.227 & - & - & - & - \\
  ST & $\widehat{\beta}_0$ & 35.097 &  0.084 & -2.884 & 2677 & 6 &  &  \\
         & $\mbox{SE}_0$ & 0.07 & 0.001 & 0.23 & - & - & - & - \\
   & RC & 0.11\% & 1.21\% & 6.21\% & 0.37\% &  &  &  \\
  \hline
\end{tabular}
\end{center}
\end{table}

In that which follows, we apply the methodology on the VB model described above to analyze a real sample of
$2687$ Cardinalfish observations. A descriptive analysis of these data are summarized in Figure~\ref{hist} and Table~\ref{AD}.
All the statistical methods considered in this study as well as the parameter estimations such as
the ECME algorithm, variance-covariance matrix computation, and diagnostic tools have been computationally implemented
in {\tt R} software (R Development Core Team, 2012) in the {\tt skewtools} package developed by Contreras-Reyes (2012).
Other methods implemented in {\tt R} by Kahm et al. (2010) consider different conditions in order to derive a conclusive
dose-response curve, for instance for a compound that potentially affects the growth curve; for example, length of the lag phase,
maximal growth rate, and stationary phase. On the other hand, {\tt skewtools} package assessing with the log-likelihood function, the
distribution of the errors, and the influential analysis of observations.

Figures~\ref{H1} and~\ref{H2} show the scatter-box-histogram plots for ages between 3 and 61 years for the studied specie, which are classified in 13
categories most of which have a length of 5 years as indicated in Table~\ref{AD}. Table~\ref{AD} shows some descriptive statistics associated with the
lengths of the species in these categories. As we can see from Figure~\ref{hist}, the empirical distribution for the lengths of younger subjects (aged
between 3 and 38 years) is symmetric to light tails. Also, this distribution has its mode in the stretch of 28-33 years, which contains 456 observations. However, the distribution associated with the lengths of older subjects presents asymmetries and heavy tails. For example, for the 43-48 year category, we observe a high frequency of 184 species with lengths between 32 and 40 cm.  Moreover, if we consider the relative frequency in Table~\ref{AD}
and the ages histogram in Figure~\ref{hist}, we again see a low number of observations for the 1-3, 48-53, 53-58 and 58-61 year age categories.

In this study we proceed to evaluate the performance of each of the models T and ST as follows (for N and SN models, the step 1 is omitted):

\begin{enumerate}
\item For the estimation of the VB model parameters from the T and/or ST models, we need to find an estimate for the  degrees of freedom parameter $\nu$ that maximizes the corresponding log-likelihood functions. To do this, Lange \& Sinsheimer (1993) recommend to compute profiles of the log-likelihood function for a given $\nu$, where the optimal $\nu$ is given for the maximum profile log-likelihood function.
\item Computed the parameter set $\widehat{\theta}$ by the ECME algorithm, we proceed to diagnose the model associated with these parameter set by the local influential analysis. A re-estimation of the parameter set, $\widehat{\theta}_{0}$ say, from the data without the influential observations is thus obtained.
\item To evaluate the change produced by the influential observations (detected by using the local influence diagnostic analysis described previously) in the estimation of each component $\beta_k$, $k=1,2,3$, of the vector of parameters $\beta$ associated  to the VB curve, with respect to the estimated parameters including the outliers, we proceed to a confirmatory analysis via the percentaged relative change (RC):
$$\mbox{RC}(\widehat{\beta}_{k0},\widehat{\beta}_k)=\left| 1-\frac{\widehat{\beta}_{0k}}{\widehat{\beta}_k} \right|\times100\%;$$
(see, e.g., Bastos et al., 2012); where $\widehat{\beta}_k$ is the estimation of $\beta_k$ from the complete data, and $\widehat{\beta}_{k0}$ is the corresponding estimation however obtained from the data without the influential observations. Therefore, the RC index represents the percentage change for a parameter when we exclude the influential observations of the data set.

\item We also compare the N, SN, T and ST models using the AIC (Akaike's information criterion) value given by Akaike (1974) defined as

$$\mbox{AIC}(\widehat{\theta})=-2[\ell(\widehat{\theta})-q]$$

associated to parameters set $\widehat{\theta}$ where $q$ is the number of parameters. 
\end{enumerate}

 Tables~\ref{CG2} and~\ref{CG3} present the fit of models T and ST, respectively, for $\nu=2,5,15,...,45$.  Table~\ref{CG4}  summarizes and compares  the T and ST model fits for $\nu=37$ and  $\nu=51$, respectively. As shown in Figure~\ref{LogL}, these values for the degrees of freedom maximize the corresponding profiles of log-likelihood functions. All of these tables include estimates of the VB parameters with their respective standard errors and the AIC and log-likelihood values for both, the full and the filtered (by influential observations) samples. Note from Figure~\ref{LogL} that considering the full sample, the T log-likelihood is maximized when $\nu=37$. Considering this value for the degrees of freedom and the sample filtered by influential observations, we obtain the estimations $\widehat{L}_{\infty}=35.095$, $\widehat{K}=0.084$ and $\widehat{t}_0=-2.943$. Results are obtained similarly for the ST model, where the ST log-likelihood is maximized when $\nu=51$. For this value of $\nu$, the estimations of the VB parameters under the filtered sample are  $\widehat{L}_{\infty}=35.097$, $\widehat{K}=0.084$ and $\widehat{t}_0=-2.884$. Consequently, we could consider the estimations obtained for $\nu=37$ as a selection threshold for the VB parameters for the T case, and the results obtained for  $\nu=51$ under the ST case. As was mentioned in Section~1, these results could be significantly modified by the incorporation of the missing data in the 1-3 age category, particularly the estimation of $t_0$ and the criteria for influence data and model selection.

 Now, we describe the main results of our analysis for the estimated parameters without influential observations. Thus, from the T model fit  with $\nu=37$ we have $\widehat{\rho}=-0\mbox{.}69$ and $\widehat{\sigma}^2=25\mbox{.}36$. Also, for the estimation of the parameters $\kappa_{1}$ and $\kappa_{2}$ associated with the $Gamma(\nu/2, \nu/2)$ mixing distribution, we obtain $\kappa_{1}=1\mbox{.}02$ and $\kappa_{2}=1\mbox{.}06$. These estimations are approximately equal to the corresponding parameter for the N and SN distributions, for which $\kappa_{1}=\kappa_{2}=1$. While from the ST model fit  with $\nu=51$, we have $\widehat{\rho}=-0\mbox{.}703$, $\widehat{\sigma}^2=34\mbox{.}87$ and $\widehat{\lambda}=0\mbox{.}39$. For the N case we also find that $\widehat{\rho}=-0\mbox{.}68$ and $\widehat{\sigma}^2=26\mbox{.}93$. Figure~\ref{K22} shows the confidence bands $\{\widehat{\eta}_t\pm z_{(1-\alpha/2)}\sqrt{Var[\widehat{\eta}_t]}\}$ obtained from the ST case, and the behavior of the VB growth curve and its respective variances represented by $\log\sqrt{Var[\widehat{\eta}_t]}$.
Here, $z_{(1-\alpha/2)}$ denote the standardized normal quartile related to a significance level given by $\alpha$ and $\eta_t=E(y_t)$ is given in (\ref{r-mean-var}).

\section{Discussion}

 With the aim of comparing the different distributions in this study, Table~\ref{CG1} shows the estimations for the VB vector of
 parameters $(L_{\infty},K, t_0)$ and its respective standard deviations obtained from the N, T, SN and ST models. These models
 are fitted using the full sample data and a sample data filtered by influential observations from the local influence criteria
 described in Section~\ref{IL}.  From the obtained  results, we can not see a perceptible difference in the estimation of
 $(L_{\infty}, K, t_0)$ through the four fitted models. However, the standard errors associated to these parameters are
 lesser for ST model. This fact occurs mainly due to the prevalence of missing observations in the range of 1 and 3 years,
 which increments $Var[\widehat{y}_t]$ related with these years. Also, in Table~\ref{CG1} we present the AIC criteria to
 compare the fit of these four models considering the filtered sample data. As was previously mentioned, the N and SN model-fits
 detect a larger quantity of influential observations (54 and 50, respectively). 
 In addition, we observe that the ST fit has a smaller value for the AIC criteria. Moreover, for this model the VB estimated parameters are $L_{\infty}=35\mbox{.}097$, $K=0\mbox{.}084$ and $t_0=-2\mbox{.}884$ considering the dataset without influential observations. Note that for the four model fits, the RC in the estimation of $L_{\infty}$ and $K$ obtained from the model fit with filter and full sample is of 0.11\% and 1.21\%, respectively;
 nevertheless, the estimation of $t_0$ presents despairs changes because the ST distribution presents the larger variation (6.21\%).

Results on a diagnostic analysis are illustrated in Figura~\ref{K3}. The Cook's diagnostic measure tends to be sensible under structural changes of the covariance matrix associated to the parameter estimators based on the T and ST models; specifically, for changes in the parameter $\nu$ (Labra et al., 2012). Note from Table~\ref{CG1} that the standard errors (SE) of the estimates obtained from the ST model tend to be smallest with respect to those obtained from the other models. In fact, we can see that the reference level $c_{\bd}$ for the T and ST cases is around to 0.004, while that for the N and SN cases this level is around to 0.002. This fact produces that, under the N or SN models, several observations are identified as influential; while under the T and ST assumptions, they are considered not influential data. Specifically, 54 (2.01\%) influential observations are identified for the N model, 50 (1.86\%) for the SN model, 7 (0.26\%) for the T model and 10 (0.37\%) for the ST model. Numerical simulations carried out by Lachos et al. (2011) showed that for similar parameter values, the T and ST distributions effectively allow a lower number of influential observations than that of the N and SN distributions.
Our results show that the optimal degrees of freedom parameters are large for the T and ST model fits. This should be interpreted as an approximation
of the T and ST model fits to the N and SN model fits, respectively. However, the ST model fit produces the estimation with the smallest standard errors and gives a minor number of influential observations, given some features of robustness of the maximum likelihood estimation under T and ST distributions (Labra et al., 2012).

The estimations of the VB parameters obtained by Cubillos et al. (2009) coincide with those obtained by G\'alvez \& Rebolledo (2011), who report  $L_{\infty}=46.8$, $K=0.147$ and $t_0=0$. However, the methodologies adopted by G\'alvez \& Rebolledo (2011) consider an exponential relationship between the otolith mass and the length, which should produce some errors related to age assign (Ojeda et al., 2010). On the other hand, we have considered observations associated to older subjects but with missing observations of younger subjects (Ojeda et al., 2010), because the younger fraction of length less than 20 cm does not appear in the catch (G\'alvez \& Rebolledo, 2011; Wiff et al., 2005). This should produce some differences in the parameter estimations with respect to that presented by Cubillos et al. (2009), who estimated a positive value for $t_0$ and an overestimation and underestimation for $K$ of the asymptotic length $L_{\infty}$, respectively. Our negative estimation of $t_0$ is produced mainly by the missing observations in the first age category (1-3], where this should be solved by the use of back-calculation (see e.g. Francis, 1990, and the references therein) or improve in techniques of otolith readings to infer its length at an earlier time or times. Consequently, considering that the values for the VB parameters have an impact on the mortality of Cardinalfish (Hewitt \& Hoenig, 2005), these differences lead to a totally different exploitation scenario.

Different distributions have been presented in this study with the aim of giving new estimation and diagnostic tools related to
methods that assume gaussian residuals in growth models. Since the normality assumption is questionable when used to analyze these observations,
the flexible class of SMSN distributions provides robust models to estimate the parameters of the VB growth curve and thus determines
appropriate mortality indexes. In addition, we have included an analysis of local influence,
which allows the identification of anomalous observations, using the $t$-Student and skew-$t$ distributions.
Given the special features of the Cardinalfish --such as longevity-- our proposal allows us to find differences
in the estimation of the VB parameters respect to the results presented in the literature. Finally,
the proposed methodology could be used to analyze the growth features of other species and for other growth models.

\section*{Acknowledgment}

The authors wish to thank to Vilma Ojeda and Rodrigo Wiff for the database used in this paper and for his helpful suggestions and comments.
Arellano-Valle's research was supported by Grant FONDECYT (Chile) $1120121$.

\section*{References}

\newenvironment{reflist}{\begin{list}{}{\itemsep 0mm \parsep 1mm
\listparindent -7mm \leftmargin 7mm} \item \ }{\end{list}}
\baselineskip 19.9pt
{\small

\refmark
Akaike, H. (1974). A new look at the statistical model identification. {\it IEEE T. Automat. Contr.}, {\bf 19}, {\bf 6}, 716-723.

\refmark
Allen, K.R. (1966). A method of fitting growth curves of the von Bertalanffy type to observed data. {\it J. Fish. Res. Board Can.}, {\bf 23}, 163-179.

\refmark
Arellano-Valle, R.B., Contreras-Reyes, J.E., Genton, M.G. (2012). Shannon Entropy and Mutual Information for Multivariate Skew-Elliptical Distributions.
{\it Scand. J. Statist.}, doi: 10.1111/j.1467-9469.2011.00774.x.

\refmark
Azzalini, A., Capitanio, A. (2003). Distributions generated by perturbation of symmetry with emphasis on a multivariate skew t distribution. {\it J. R. Stat.
Soc. Ser. B}, {\bf 65}, {\bf 2}, 367-389.

\refmark
Basso, R.M., Lachos, V.H., Barbosa, C.R., Ghosh, P. (2010). Robust mixture modeling based on scale mixtures of skew-normal distributions. {\it Comput.
Stat. Data An.}, {\bf 54}, {\bf 12}, 2926-2941.

\refmark
Bastos, L.G., Nobre, J.S., Freitas, S.M. (2012). On linear mixed models and their influence diagnostics applied to an actuarial problem.
{\it Chil. J. Stat.}, {\bf 3}, {\bf 1}, 57-73.


\refmark
Branco, M., Dey, D. (2001). A general class of multivariate skew-elliptical distribution. {\it J. Multivariate Anal.}, {\bf 79}, {\bf 1}, 93-113.


\refmark
Cailliet, G.M., Andrews, H.A. (2008). Age-validated longevity of fishes: Its importance for
sustainable fisheries. {\it Fisheries for global welfare and environment}, 5th congress 2008, 103-120.

%

\refmark
Contreras-Reyes, J.E. (2012). {\it R package skewtools: Tools for analyze Skew-Elliptical distributions and related models (version 0.1.1)}. URL\\ \url{http://cran.rproject.org/web/packages/skewtools}


\refmark
Cook, R.D. (1986). Assessment of local influence (with discussion). {\it J. R. Stat. Soc. Ser. B}, {\bf 48}, {\bf 2}, 133-169.

\refmark
Cope, J.M., Punt, A.E. (2007). Admitting ageing error when fitting growth curves: an example using the von Bertalanffy growth function with random
effects. {\it Can. J. Fish. Aquat. Sci.}, {\bf 64}, 205-218.

\refmark
Cubillos, L.A., Aguayo, M., Neira, M., Sanhueza, E., Castillo-Jord\'an, C. (2009). Age verification and growth of the Chilean cardinalfish Epigonus
Crassicaudus (de Buen, 1959) admitting ageing error. {\it Rev. Biol. Mar. Oceanogr.}, {\bf 44}, {\bf 2}, 417-427.

\refmark
Francis, R. (1990). Back-calculation of fish length: a critical review. {\it J. Fish Biol.}, {\bf 36}, 883-902.

\refmark
G\'alvez, M., Rebolledo, H. (2001). Length-Composition and length-weight relationship in cardinalfish (Epigonus crassicaudus) of the central-southern are
off Chile. {\it Invest. Mar., Valpara\'iso}, {\bf 29}, {\bf 2}, 39-49.

\refmark
G\'alvez, M., Rebolledo, H., Pino, C., Cubillos, L., Sep\'ulveda, A. \& Rojas, A. (2000). {\it Par\'ametros biol\'ogico-pesqueros y evaluaci\'on del stock de besugo (Epigonus crassicaudus)}. Informe Final, Instituto de Investigaci\'on Pesquera, VIII Región, Talcahuano, 110 pp.

\refmark
Gamito, S. (1998). Growth models and their use in ecological modelling: an application to a fish population. {\it Ecol. Model.}, {\bf 113}, 83-94.

\refmark
Hewitt, D.A., Hoenig, J.M. (2005). Comparison of two approaches for estimating natural mortality based on longevity. {\it Fish. Bull.}, {\bf 103}, 433-437.

\refmark
Kahm, M., Hasenbrink, G., Lichtenberg-Frat\'e, H., Ludwig, J., Kschischo, M. (2010). grofit: Fitting Biological Growth Curves with R. {\it J. Stat. Softw.}, {\bf 33}, {\bf 7}, 1-21.

\refmark
Kimura, D.K. (1980). Likelihood methods for the von Bertalanffy growth curve. {\it Fish. Bull.}, {\bf 77}, {\bf 4}, 765-776.

\refmark
Kimura, D.K. (1990). Testing nonlinear regression parameters under heteroscedastic, normally distributed errors. {\it Biometrics}, {\bf 46}, {\bf 3}, 697-708.

\refmark
Labra, F.V., Garay, A.M., Lachos, V.H., Ortega, E.M.M. (2012). Estimation and diagnostics for heteroscedastic nonlinear regression models based
on scale mixtures of skew-normal distributions. {\it J. Statist. Plann. Inference}, {\bf 142}, {\bf 7}, 2149-2165.

\refmark
Lachos, V.H., Ghosh, P., Arellano-Valle, R.B. (2010). Likelihood based inference for skew-normal independent
linear mixed models. {\it Stat. Sinica}, {\bf 20}, 303-322.

\refmark
Lachos, V.H., Bandyopadhyay, D., Garay, A.M. (2011). Heteroscedastic nonlinear regression models based on scale mixtures of skew-normal distributions.
{\it Stat. Probabil. Lett.}, {\bf 81}, {\bf 8}, 1208-1217.

\refmark
Lange, K.L., Sinsheimer, J.S. (1993). Normal/independent distributions and their applications in robust regression. {\it J. Comput. Graph. Stat.}, {\bf 2}, {\bf 2}, 175-198.

\refmark
Liu, C., Rubin, D.B. (1994). The ECME algorithm: A simple extension of EM and ECM with faster monotone convergence. {\it Biometrika}, {\bf 81}, {\bf 4}, 267-278.

\refmark
Ojeda, V., Wiff, R., Labr\'in, C., Contreras, F. (2010). Longevity of cardinalfish Epigonus crassicaudus in Chile: is it similar to that of its relatives? {\it Rev. Biol. Mar. Oceanogr.}, {\bf 45}, {\bf 3}, 507-511.

\refmark
Poon, W., Poon, Y.S. (1999). Conformal normal curvature and assessment of local influence. {\it J. R. Stat. Soc. Ser. B}, {\bf 61}, {\bf 1}, 51-61.


\refmark
R Development Core Team (2012). {\it R: A Language and Environment for Statistical Computing}, R Foundation for Statistical Computing, Vienna, Austria. ISBN
3-900051-07-0, URL \url{http://www.R-project.org}

\refmark
von Bertalanffy, L. (1938). A quantitative theory of organic growth (Inquiries on growth laws. II). {\it Hum. Biol.}, {\bf 10}, 181-213.

\refmark
Wiff, R., Quiroz, J.C., Tascheri, R. (2005). Exploitation status of cardinalfish (Epigonus crassicaudus) in Chile. {\it Invest. Mar., Valpara\'iso}, {\bf 33}, {\bf 1}, 57-67.

}

\end{document}